%% file: 1_Main.tex
\documentclass[12pt,a4paper]{article}
\usepackage[utf8]{inputenc}
\usepackage{graphicx} 
\usepackage{authblk}
\usepackage{float}
\usepackage{gensymb}
\usepackage{comment}
\usepackage{upgreek}
\usepackage{hyperref}
\usepackage{ragged2e} 
\usepackage[dvipsnames]{xcolor} 
\usepackage{amssymb}
\usepackage{multirow}
\usepackage[labelfont=bf]{caption}
\usepackage[labelsep=period]{caption}

\usepackage{setspace}
\usepackage{lineno}

\usepackage{geometry}
\usepackage[superscript,biblabel]{cite}

\title{Ferrofluidic Aqueous Two-Phase System with Ultralow Interfacial Tension, Instabilities and Pattern Formation}
\author[1]{Carlo Rigoni*}
\author[1]{Bent Harnist}
\author[1]{Grégory Beaune}
\author[1]{Jaakko V. I. Timonen*}
\affil[1]{Department of Applied Physics, Aalto University School of Science, Puumiehenkuja 2, 02150 Espoo, Finland}
\date{}

\begin{document}

\maketitle

\section*{Abstract}
\textbf{Ferrofluids are strongly magnetic fluids consisting of magnetic nanoparticles dispersed in a carrier fluid. Besides their technological applications, they have a tendency to form beautiful and intriguing patterns when subjected to external static and dynamic magnetic fields. Most of the patterns occur in systems consisting of two fluids: one ferrofluidic and one non-magnetic (oil, air, etc.), wherein the fluid-fluid interface deforms as a response to magnetic fields. Usually, the fluids are completely immiscible and so the interfacial energy in this systems is very large. Here we show that it is possible to design a fully aqueous ferrofluid system by using phase separation of incompatible polymers. This continuous aqueous system allows an ultralow interfacial tension (down to 1 \(\upmu\)N/m) and nearly vanishing pinning at three phase contact lines. We demonstrate the normal-field instability with the system and focus on the miniaturization of the pattern length from the typical \(\sim\)10 mm size down to \(\sim\)200 \(\upmu\)m. The normal-field instability is characterized in glass capillaries of thickness comparable to the pattern length. This system paves way towards interesting physics such as the interaction between magnetic instabilities and thermal capillary waves and offers a way to evaluate extremely small interfacial tensions.}

\clearpage
\justify
Ferrofluidic systems can exhibit a great variety of properties and functionalities. From the possibility to control the size and composition of the nanoparticles (NPs), to the extremely wide choice of different stabilizing agents, or the fact that the carrier liquid can be chosen among both polar and non-polar liquids, they show a flexibility that makes them particularly useful in a wide range of applications. \cite{zhang2019flexible} The other main reason for their success lies in the particular phenomena that arise when they interact with the magnetic field: from the magnetic hyperthermia driven by oscillating fields,\cite{deatsch2014heating,rosensweig2002heating} to the magnetoviscous effect caused by the strong interaction between NPs, \cite{odenbach2002magnetoviscous} to the magnetic instabilities generated at the interface between two phases with different magnetizations. \cite{rosensweig2013} In particular, the magnetic instabilities include a large number of phenomena related to pattern formation with the most significant examples being the normal-field or Rosensweig and the labyrinthine instabilities \cite{cowley1967interfacial,rosensweig1983labyrinthine,dickstein1993labyrinthine}. Other noteworthy examples are the interface deformations of droplets, \cite{afkhami2010deformation,rigoni2016static} droplet splitting and self-assembly, \cite{rigoni2018division,timonen2013switchable} and soliton like features at the interface. \cite{richter2005two}
The normal-field instability has proven to be a universal phenomenon since it has also been observed in very exotic systems like the magnetic Bose-Einstein condensates and dissipative electrodriven gradients of NPs.\cite{kadau2016observing, cherian2020Electroferrofluids} In fluid environments, the normal-field instability has been studied almost exclusively in systems relying on one magnetic phase (ferrofluid) with an interface towards a non-magnetic phase (air, non-magnetic liquid, etc.).\cite{cowley1967interfacial,gollwitzer2006surface,richter2009surface} In these systems the periodicity of the normal-field instability patterns is determined by the interfacial tension between the two phases through:\cite{rosensweig2013}
\begin{equation}
\lambda_0 = 2 \pi \sqrt{\frac{\gamma}{g \Delta \rho}}
\label{equation1}
\end{equation}
where \(\lambda_0\) is the pattern periodicity, \(\gamma\) is the interfacial tension, \(g\) is the gravitational acceleration and \(\Delta \rho\) is the density difference between the two phases. The interfacial tension also influences the critical value of magnetization to be reached for the onset of the instability:\cite{rosensweig2013}
\begin{equation}
      \Delta M^2_c = \frac{2}{\mu} \left(1+ \frac{\mu}{\mu'}\right) \sqrt{g \Delta\rho\gamma}
      \label{equation2}
\end{equation}
where \(\Delta M_c\) is the magnetization difference between the two phases at the threshold at which the pattern appears and \(\mu\) and \(\mu'\) are the magnetic permeabilities of the ferrofluid and the non-magnetic phase, respectively.
Therefore, classic systems have long periodicities because of the high interfacial tension and, for the same reason, the value of \(\Delta M_c\) is usually high: this limits the possibilities of observation of this phenomenon to very concentrated systems. It would be interesting to go smaller and miniaturize the pattern lenght below the typical mm size to probe the interaction between the pattern itself and other microscale behaviors like the thermal capillary waves. \cite{aarts2004direct} Achieving this, is unfortunately not feasible in most cases since the interfacial tension of water with air is around 72 mN/m, oils around 30 mN/m and even fluorinated liquids 10-20 mN/m at room temperature. Two liquid phases seem more promising since it is possible to go down to 0.1 mN/m tuning the strenght of the interface with proper surfactants but even smaller values would be desired. \cite{latikka2020ferrofluid} \\
In this article, we demonstrate a ferrofluidic system where we can go several orders of magnitude smaller in the interfacial tensions than the classical systems. The presented system is based on the so called Aqueous Two Phase Systems (ATPS). These are obtained by mixing aqueous solutions of incompatible polymers and/or salts which generate a phase separation in the water solution with each phase having a higher concentration of one of the two components. It is important to notice that both phases are fully aqueous with an interfacial tension between them of orders of magnitude lower than the one between two liquids systems around \(\sim 1 \upmu\)N/m.\cite{atefi2014ultralow} 
The interfacial tension can also be regulated by controlling the polymers (salts) concentrations: increasing the concentration will strengthen the interface while diluting the solution will weaken it. Concentrating/diluting the solution also allow to control the partitioning of macromolecules and colloidal particles in these systems. For this reason ATPS have been historically used in a wide range of applications to obtain selective partitioning of biomolecules and nanoparticles exploiting the different physico-chemical affinities for the two phases. \cite{walter1985partitioning,soohoo2009microfluidic,mastiani2019polymer,bai2020all, pereira2020aqueous} \\

\begin{figure}[H]
\centering
\includegraphics{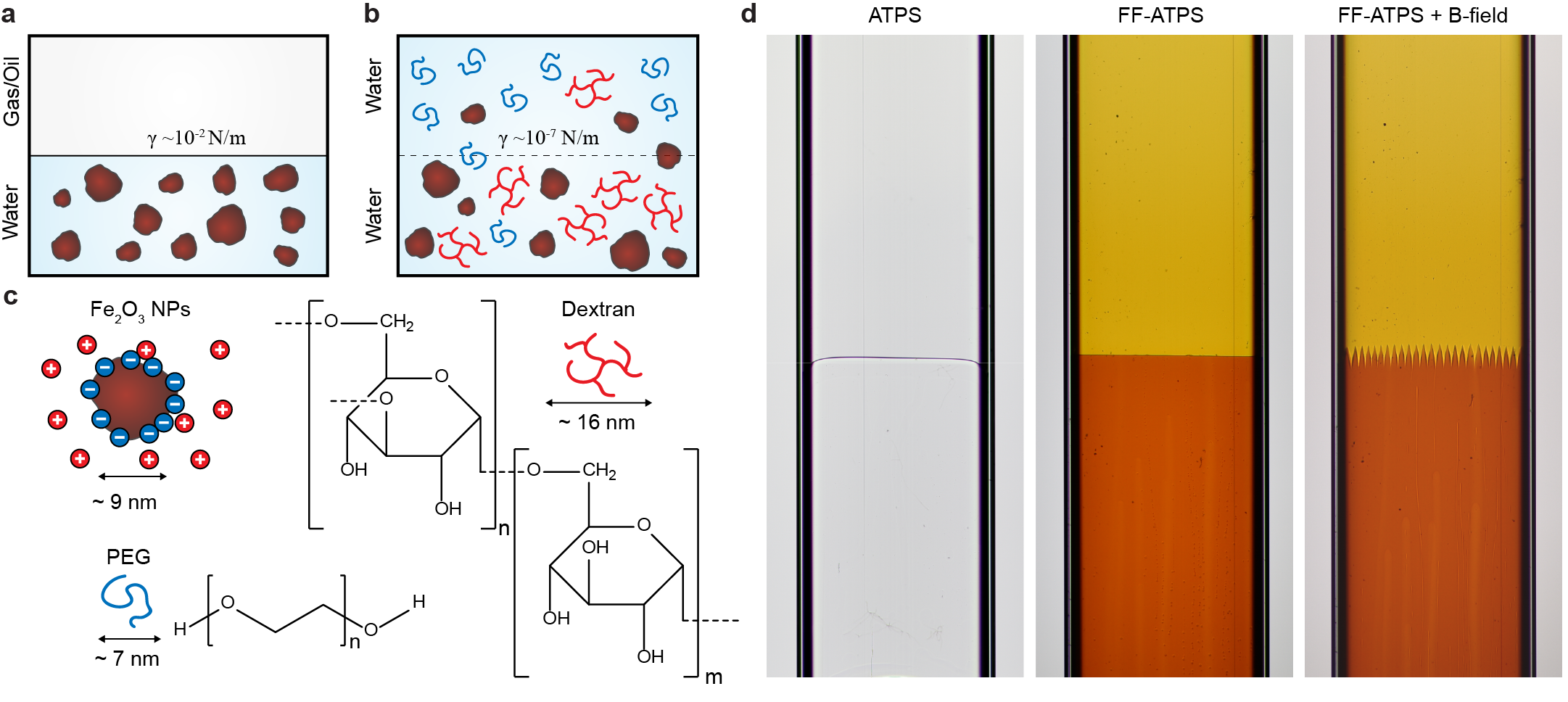}
\caption{\textbf{Creating FF-ATPS and proving their magnetic field responsiveness.}
\textbf{a,} Scheme of the classical ferrofluidic system. \textbf{b,} Scheme of the FF-ATPS. \textbf{c,} Scheme of the system composition. \textbf{d,} Photographs of a standard PEG-Dextran aqueous two phase system (left), a PEG-Dextran FF-ATPS (center) and the same PEG-Dextran FF-ATPS under the influence of a vertical magnetic field of 20.0 mT (right). All images are relative to each sample in a vertical glass capillary with rectangular inner shape (size 0.20\(\times\)4.00 mm). }
\label{Fig1}
\end{figure}

In the present work, we use an ATPS system based on polyethylene glycol (PEG, Mw = 34 kDa) and dextran (Mw = 485 kDa) above a certain critical concentration (as in Atefi 2014 \cite{atefi2014ultralow}). To this system we add sodium citrate stabilized maghemite NPs that mainly partition to the dextran-rich phase (Fig. \ref{Fig1}c, d, see the Methods section and the Supplementary Note S2 for more details on the composition of the system). After mixing, the partitioned maghemite NPs still maintain a good dispersion stability and so all the typical properties of the ferrofluidic systems are preserved thus creating what we call a Ferrofluidic Aqueous Two Phase System (FF-ATPS). The main difference between this new system and the classic ferrofluidic systems is linked to the nature of the interface: in the classical ferrofluidic systems the NPs are present only in the magnetic phase and the two liquids are chemically different and completely immiscible (Fig. \ref{Fig1}a), in the FF-ATPS the interface is much weaker and the difference in composition is only quantitative with the NPs and polymers that can even cross the interface (Fig. \ref{Fig1}b). Despite the large differences between the two systems, it is still possible to take advantage of the partitioning of the magnetic NPs to observe magnetic instabilities at the interface (Fig. \ref{Fig1}d). In the following discussion we demonstrate how to take advantage of the described system to tune the interfacial tension and the partition of NPs with the aim to study the onset of the well known normal-field instability in different conditions and geometrical configurations. \\

\begin{figure}[H]
\centering
\includegraphics{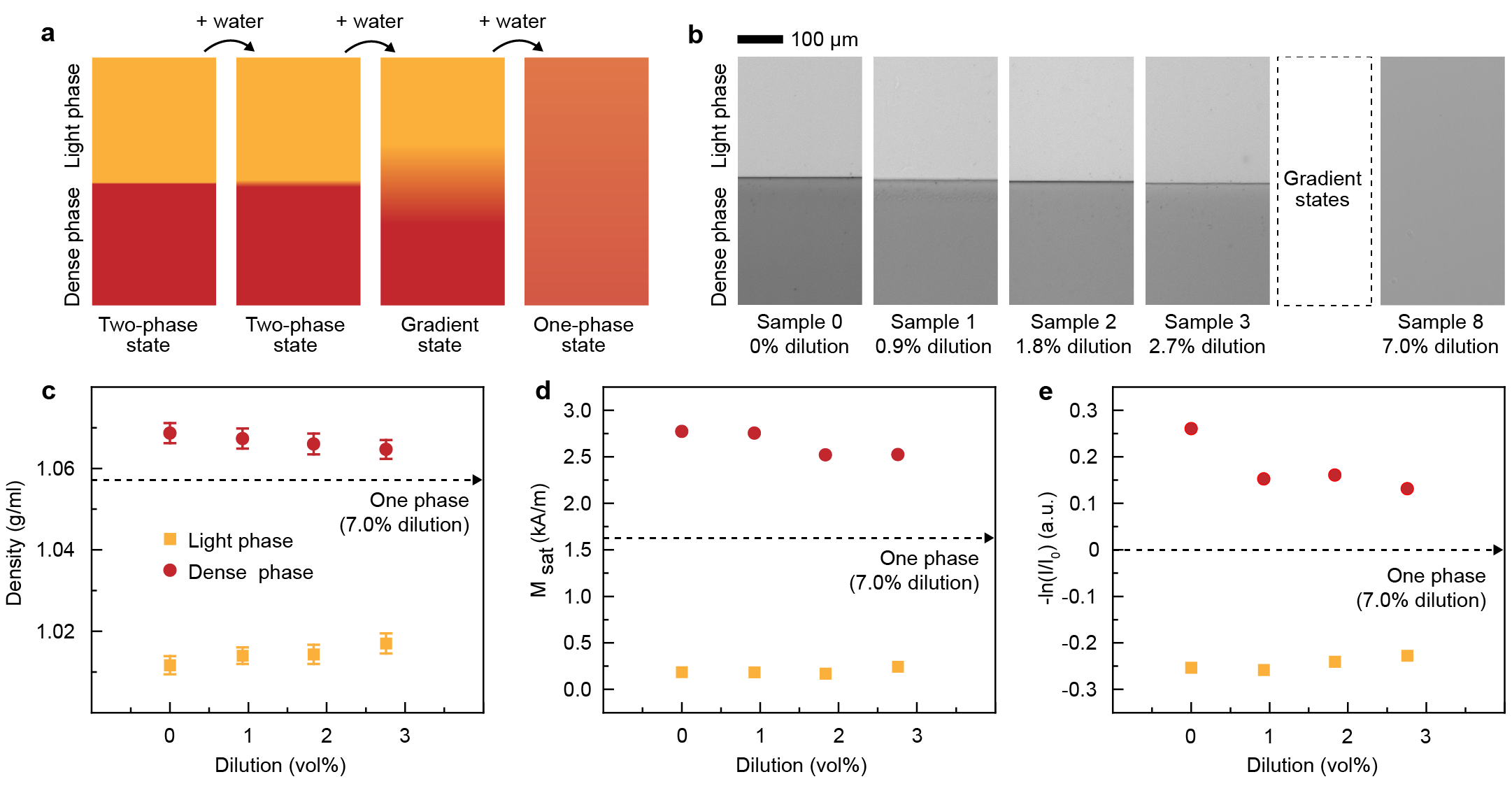}
\caption{\textbf{Characteristics of the prepared FF-ATPS samples.}
\textbf{a,} Scheme of the behavior of an FF-ATPS after progressive dilution with water. \textbf{b,} Images of the interface (Samples 0, 1, 2, 3) or the central region (Sample 8) of the different prepared samples diluting sample 0. All images are relative to each sample in a vertical glass capillary with rectangular inner shape (size 0.10\(\times\)2.00 mm). \textbf{c,} Density measurements results for the dense and light phase for all the prepared sample. \textbf{d,} Saturation magnetization measurements results for the dense and light phase for all the prepared sample (see Extended Data Fig. 2 for further details). \textbf{e,} \(-\ln(I/I_0)\) measurements results for the dense and light phase for all the prepared sample, the intensity of the single phase \(I_0\) has been taken as reference. }
\label{Fig2}
\end{figure}

Progressively diluting a FF-ATPS sample weakens the interface by lowering the interfacial tension between the two phases until a transition point at which the interface dissolves into a concentration gradient of NPs (Fig. \ref{Fig2}a). If this state is further diluted a uniform state is reached. Following this approach we diluted a concentrated FF-ATPS (PEG: 1.80\%, Dextran 3.93\%, NPs 3.15\%, see Supplementary Note S2) dispersion with controlled amounts of water to obtain four different samples with different interfacial tensions and a fifth sample with uniform NPs concentration (Fig. \ref{Fig2}b). The samples were coded as 0, 1, 2, 3 and 8 representing the approximate water added in vol\%.
A measurement of the interfacial tension with the standard techniques (pendant drop, sessile drop) is particularly difficult because of the low interfacial tension and because the ferrofluid is opaque to visible light at the thicknesses needed for those measurements. Very approximate measurements with puddle area evaluation gave results going from \(\sim\)4 \(\upmu\)N/m for sample 0 to \(\sim\)0.1 \(\upmu\)N/m for sample 3 (for more information see the Methods section). The partitioning of the NPs in the system instead can be quantified with at least three different methods. We measured the density and the saturation magnetization (SM) and evaluated the concentration of NPs in all five samples (Fig. \ref{Fig2}c-e) in both the NPs rich-phase (\textit{Dense phase}) and the NPs poor-phase (\textit{Light phase}). We can evaluate the different concentrations in the samples by looking at the light absorption and using the Beer-Lambert law to obtain:
\begin{equation}
    \Delta n = - \frac{1}{\alpha} \ln \left( \frac{I}{I_0}\right)
    \label{equation3}
\end{equation}
where \(\Delta n\) is the difference in NPs concentration between two semitransparent liquids layers of the same thickness with intensity of the light passing through \(I\) and \(I_0\) respectively (see more for the derivation of the formula in Supplementary Note S6). The parameter \(\alpha\) can not be exactly calculated in this case but evaluating the value of \(-\ln ({I}/{I_0})\) still gives us a good insight in the evolution of the concentration of NPs while diluting the system. In this case we used the intensity of the uniform phase state of sample 8 as reference \(I_0\). All the different methods confirm that the progressive dilution of the main sample leads to a small reduction of partitioning of the NPs inducing a decrease of the density/SM/concentration in the dense phase and a parallel increase of the same properties in the light phase. The trend showed by these measurements converges to the obtained value for the one phase sample that have properties in the middle between the dense and the light phase. The fact that the interfacial tension changes a lot with dilution but NP partitioning does not change as much (as proven by almost no change in density, MS and transmittance) suggests that NPs strongly prefer the dextran-rich phase, irrespective of the amount of dextran molecules present in this phase compared to PEG.  Even small extra dextran makes NPs go into that phase almost completely (as shown by SM of PEG phase being almost zero all the time).

\begin{figure}[H]
\centering
\includegraphics{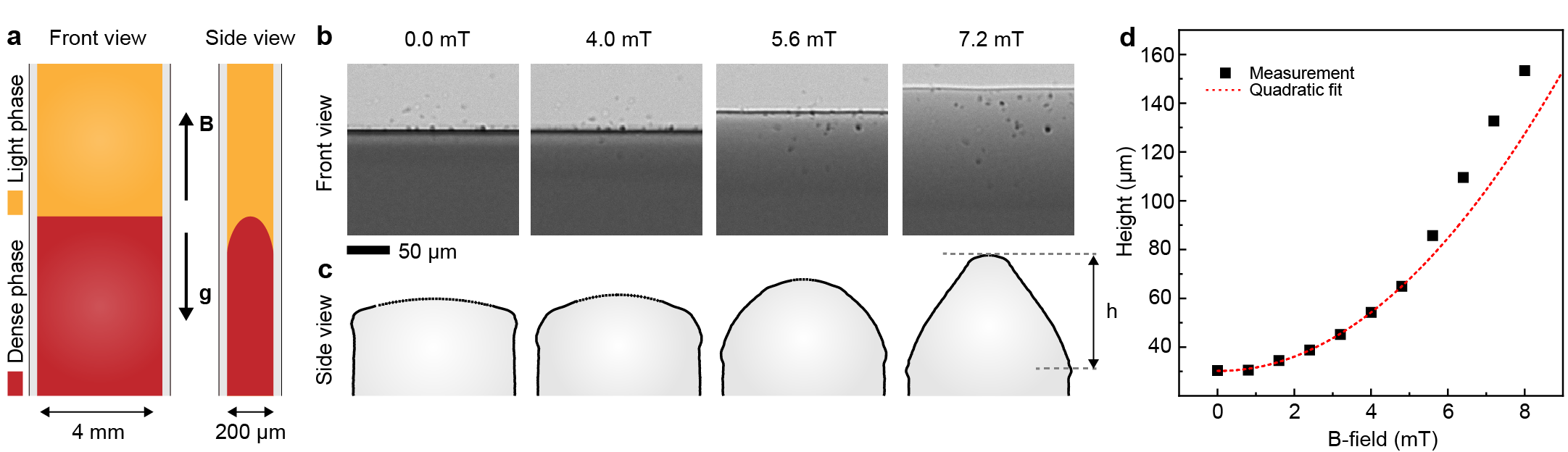}
\caption{\textbf{Meniscus elongation inside a vertical glass capillary in a FF-ATPS.}
\textbf{a,}  Scheme of the  geometry of the glass capillary and how the magnetic field and gravity are directed relative to it and relative to the interface between the light and dense phase. \textbf{b,} Images of the front view of the capillary at different magnetic field strengths. All images are relative to sample 0 in a vertical glass capillary with rectangular inner shape (size 0.20\(\times\)4.00 mm). \textbf{c,} Reconstruction of the section profiles along the y-axis in function of the magnetic field applied. \textbf{d,} Graph of the height of the meniscus in function of the magnetic field applied. }
\label{Fig3}
\end{figure}

To observe the behavior of the obtained FF-ATPS in magnetic field we used capillary glasses of different sizes. The main reason behind this is to be able to use optical techniques to visualize the patterns as thin layers of the samples are still partially transparent to visible light. If we consider a vertical glass capillary (Fig. \ref{Fig3}a), a first phenomenon that can be described is the elongation of the interface along  the applied magnetic field. Increasing the strength of the magnetic field induces a progressive larger elongation of the meniscus (Fig. \ref{Fig3}b). The cross section shape of the meniscus can be reconstructed from the front images by taking advantage of the Beer-Lambert law for light attenuation: 
\begin{equation}
    h (z) = - \frac{1}{\alpha} \ln \left( \frac{I(z)}{I_{ref}}\right)
    \label{equation4}
\end{equation}
where \( h (z)\) is the thickness of the dense phase in function of the z coordinate, \(\alpha\) is a parameter that includes the attenuation factor in both the dense and light phase, \(I(z)\) is the intensity of light in function of the z coordinate and \(I_{ref}\) is the intensity in the region where the light phase occupies the whole thickness of the capillary (see more for the derivation of the formula in Supplementary Note S6). In this case, knowing the thickness of the capillary (which in the particular case of the analysis for Fig. \ref{Fig3} is 200 \(\upmu\)m) allows us to calibrate the unknown parameter \(\alpha\) in the Beer-Lambert law and so to extract the exact thickness values of the interface from light attenuation data (see Extended Data Fig. 3). In the very proximity of the tip of the meniscus the Beer-Lambert law cannot be used because the light is strongly refracted by the interface and so the intensity is not proportional to the concentration and thickness. To solve the problem we performed a parabolic fitting between the two mirrored parts of the thickness profile and so obtained a full reconstruction of the cross-section shape of the meniscus (Fig. \ref{Fig3}c). As can be noticed, the meniscus is originally almost flat and then curves first to a hemispherical and then to a conical section. The shape at zero field is determined by the value of the capillary length \(\lambda = \sqrt{\gamma/\Delta\rho g}\) that in this case is much smaller that the thickness of the capillary (\(\lambda \sim\) 80 \(\upmu\)m with \(\gamma \sim 4\) \(\upmu\)N/m and \(\Delta\rho \sim 60\) kg/m\(^3\), see the Methods section for the interfacial tension and density determination). It is possible to measure the change of height of the meniscus in function of the magnetic field applied by evaluating the distance between the base and the tip of the meniscus (Fig. \ref{Fig3}d). This variation exhibits a clear quadratic dependence at low field while the deformation is small. The deviation from the quadratic dependence can be associated to the transition from the hemispherical to the conical section of the meniscus. 


\begin{figure}[H]
\centering
\includegraphics{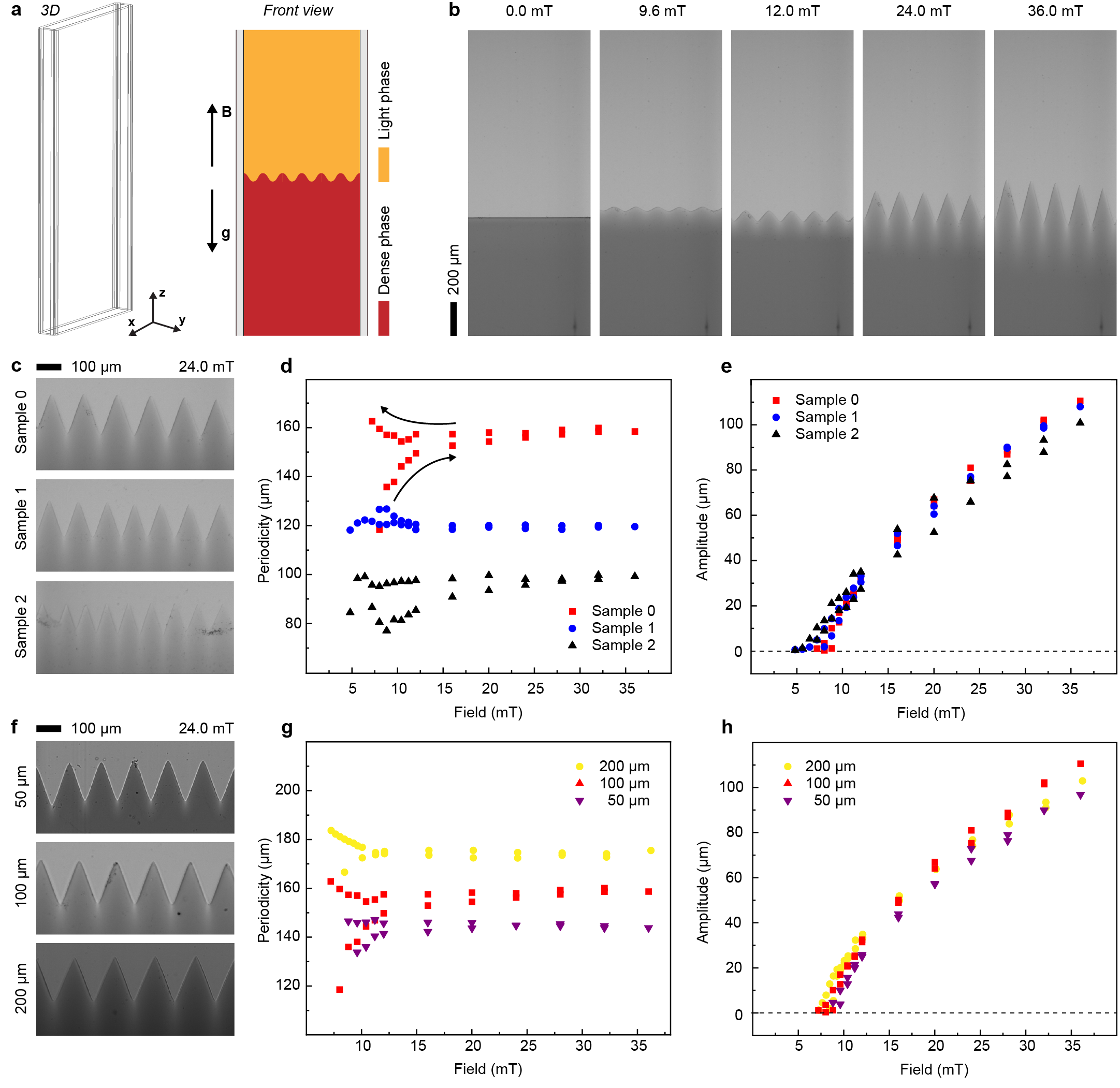}
\caption{\textbf{Meniscus magnetic normal-field instability of a FF-ATPS in a vertical glass capillary.}
\textbf{a,} Scheme of the  geometry of the glass capillary and how the magnetic field and gravity are directed relative to it and relative to the interface between the light and dense phase. \textbf{b,} Images of the front view of the capillary at different magnetic field strength. All images are relative to sample 0 in a vertical glass capillary with rectangular inner shape (size 0.10\(\times\)2.00 mm). \textbf{c,} Images of the front view of the capillary for three different samples at 24.0 mT field strength. All images are relative to samples in a vertical glass capillary with rectangular inner shape (size 0.10\(\times\)2.00 mm). \textbf{d,} Graph of the periodicity of the pattern in function of the magnetic field applied for three different samples. \textbf{e,} Graph of the amplitude of the pattern in function of the magnetic field applied for three different samples. \textbf{f,} Images of the front view of the capillary for sample 0 at 24.0 mT field strength in three different glass capillaries. \textbf{g,} Graph of the periodicity of the pattern in function of the magnetic field applied for three sizes of the capillary. \textbf{h,} Graph of the amplitude of the pattern in function of the magnetic field applied for three sizes of the capillary.     }
\label{Fig4}
\end{figure}

Further increasing the magnetic field in the case of the elongated meniscus in a vertical capillary induces a magnetic normal-field instability (Fig. \ref{Fig4}a). At first the instability appears as a sinusoidal pattern on top of the deformed interface along the x direction. If the magnetic field is further increased the amplitude of the pattern increases and the pattern slowly transitions from a sinusoidal profile to a triangular wave one with sharp spikes and valleys (Fig. \ref{Fig4}b). 
We first studied the properties of the magnetic instability pattern in the case of different dilutions of the FF-ATPS system in a glass capillary of rectangular section of 0.10 mm thickness (Fig.  4c-e). The first thing that can be noticed is that diluting the system induces an average decrease in the pattern periodicity from the \(\sim\) 160 \(\upmu\)m of sample 0 to the \(\sim\) 80 \(\upmu\)m of sample 3 (Fig.  4d). 
Furthermore, there is a small dependence of the pattern periodicity from the magnetic field and there is a hysteresis behavior linked to it with the pattern periodicity at the end of a measurement cycle being larger than the initial one observed while increasing the field. This behavior can be observed also in fully 3D systems of pools of ferrofluid. \cite{gollwitzer2006surface} Regarding the amplitude of the pattern we can observe that the three samples behaved very similarly increasing the height of the peaks non-linearly in function of the magnetic field applied (Fig. \ref{Fig4}e, Extended Data Fig. 5). The exact B-field value for the onset of the instability can be extracted by linear fit of the data for small deformation of the interface (see Extended Data Fig. 4). The value of the critical threshold field decreases from 7.6 \(\pm\) 0.9 mT for sample 0 with the highest interfacial tension to 5.3 \(\pm\) 0.9 for sample 2 with the lowest interfacial tension. This kind of pattern resembles previous studies of the normal-field instability in 2D systems \cite{flament1996measurements} with the main difference being that in our case the system has the additional thickness of the capillary in which it can deform. Despite this, we can try in first approximation to use the model for the 2D systems in our case and calculate an approximate value of the interfacial tension from it. We can write a relationship that links the values of the critical magnetization \(\Delta M_c\) and the critical wave number \(k_c\) as: \cite{flament1996measurements}
\begin{equation}
\frac{\mu_0 \Delta M_c^2 h}{2 \gamma} = \frac{(k_ch)^2+(k_0h)^2}{\int_0^{+\infty} dt \frac{1+\cos(t)}{t^2}\left( \sqrt{(k_ch)^2+t^2}-t\right)} = \frac{(k_ch)^2+(k_0h)^2}{f(k_ch)}
    \label{equation5}
\end{equation}
where \(\mu_0\) is the vacuum magnetic permeability, \(h\) is the thickness of the capillary and \(k_0\) is the critical wave number of the normal-field instability in a pool of ferrofluid. Now \(k_0 = 2\pi/\lambda_0\) where \(\lambda_0\) is the critical pattern periodicity calculated as in eq. \ref{equation1}. This allow us to rewrite eq. \ref{equation5} to calculate the interfacial tension as:
\begin{equation}
\gamma =  \frac{1}{(k_ch)^2} \cdot \left[\frac{\mu_0 \Delta M_c^2 h }{\pi} \cdot f(k_ch) -h^2\Delta\rho g\right] 
    \label{equation6}
\end{equation}
We can now use the values of the critical B-field measured with the fit from the amplitude graph and the values of the periodicity near the critical point to calculate the interfacial tension. The pattern periodicity near the critical point is measured to be: 135.7 \(\pm\) 0.5 \(\upmu\)m for sample 0, 118.2 \(\pm\) 0.5 \(\upmu\)m for sample 1 and 86.5 \(\pm\) 0.5 \(\upmu\)m for sample 2. This leads to the following approximate values for \(\gamma\): \(\sim\)1200 nN/m for sample 0, \(\sim\)700 nN/m for sample 1 and \(\sim\)330 nN/m for sample 2 (see Supplementary Note S4 for further details on the calculations). These values are only an approximation since already in the original paper the authors warned about the several possible discrepancies of the model due to the geometry or the effect of the demagnetizing field \cite{flament1996measurements}. However the model is precise enough to evaluate the order of magnitude of the interfacial tension and the decreasing trend of it in function of the dilution of the system. We then studied the behavior of a single FF-ATPS sample in glass capillaries of different thicknesses following the same protocol for the magnetic field as in the previous measurements (Fig. 4f). Consistently with the 2D model for the instability, the pattern periodicity seems to be linked to the thickness of the capillary (Fig. 4g). The differences between the sample behaviors in this case is even less evident in the amplitude data (Fig. \ref{Fig4}h, Extended Data Fig. 6). Furthermore, the critical B-field appears to be the same in the three capillaries (see Extended Data Fig. 4 and 6).

\begin{figure}[H]
\centering
\includegraphics{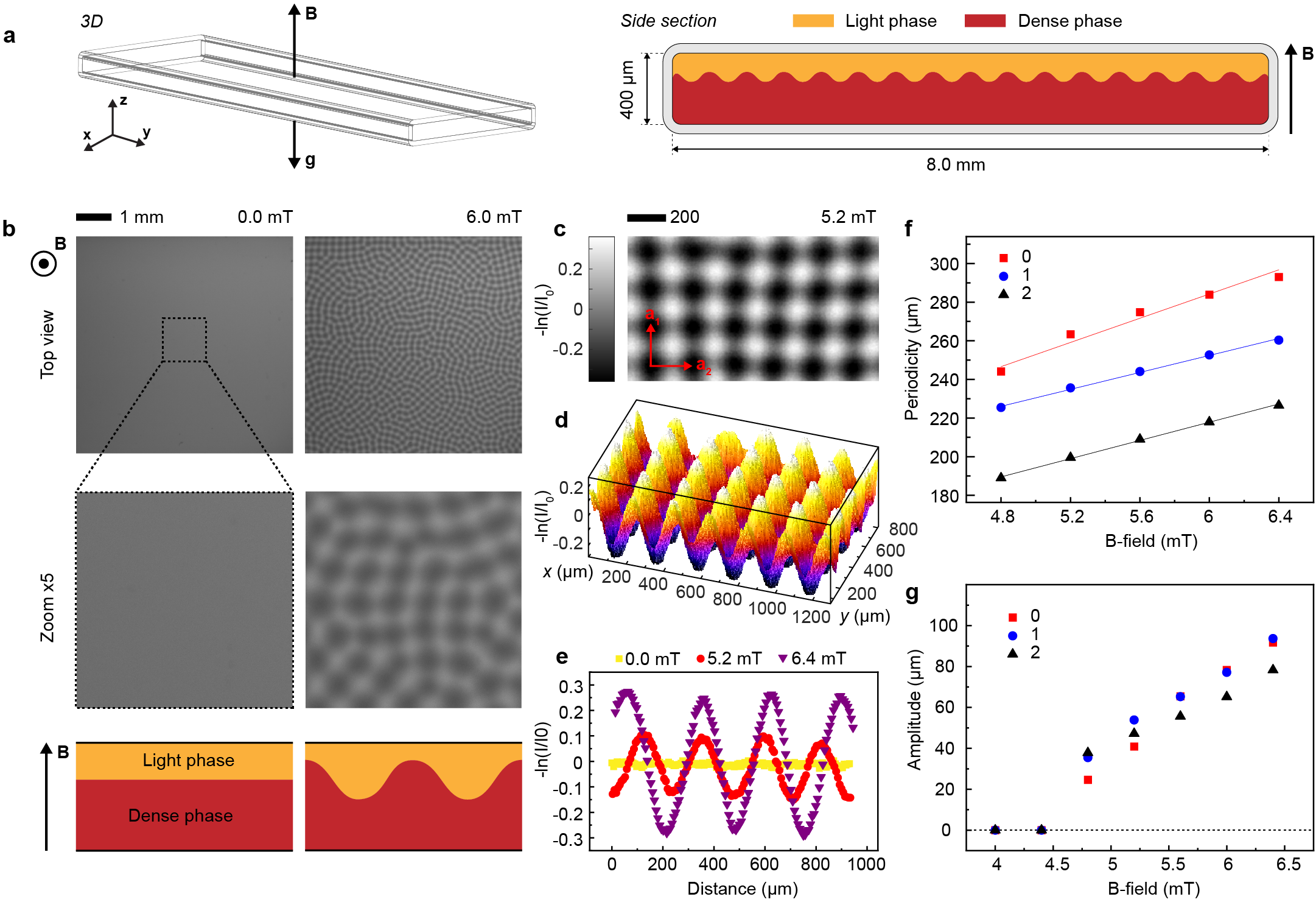}
\caption{\textbf{Interfacial normal-field instability of a FF-ATPS in an horizontal glass capillary.}
\textbf{a,}  Scheme of the  geometry of the glass capillary and how the magnetic field and gravity are directed relative to it and scheme of the  section along the x-z plane of the glass capillary and how the interface between the two phases deforms. 
\textbf{b,} Images of the view along the z-axis in function of the magnetic field applied (first row), zoom of the central area of the images (second row) and scheme of the proposed shape of the interface between the two phases for a section along the x-z plane. 
\textbf{c,} \(-\ln(I/I_0)\) image detail of the pattern, the two red arrows represent the two main directions in the unit cell of the pattern lattice.
\textbf{d,} 3-dimensional reconstruction of the shape of the pattern.
\textbf{e,} Graph of the \(-\ln(I/I_0)\) value calculated along the \textbf{a}\(_1 + \)\textbf{a}\(_2\) direction.
\textbf{f,} Graph of the periodicity of the pattern in function of the magnetic field applied for three different samples. 
\textbf{g,} Graph of the amplitude of the pattern in function of the magnetic field applied for three different samples.    }
\label{Fig5}
\end{figure}

To complete the description of the system we considered the case of an horizontal capillary (Fig. 5a). For small deformations of the interface this configuration is very similar to the classical pool of ferrofluid in which the normal-field instability as been observed first. \cite{cowley1967interfacial}
The overall behavior in function of the magnetic field in this case is much more rich due to the vertical confinement that limits the space in which the magnetic instabilities can develop. In this configuration is important to determine how much light and dense phase there are in the capillary to evaluate the thicknesses of the two layers. In all the samples we have prepared the dense phase is roughly three times more abundant than the light phase (for further details on the determination of the content of the capillaries see the Extended Data Fig. 7d and the Methods section). While increasing the field step by step the first thing we can observe is that, once reached a threshold value of magnetic field, a magnetic field instability similar to the normal-field instability occurs at the interface changing the local thickness of the dense phase and so creating a pattern characterized by darker and brighter spots (Fig. \ref{Fig5}b, second column). In this state the perturbation of the interface is small and does not touch neither the upper or the lower glass wall. The main difference between this pattern and the typical normal-field instability pattern is that in this case it does not have a long range order: the consistency in the alignment of the peaks is confined in small domains. Inside the domains the square pattern seems to be favourite instead of hexagonal one typical of the classical normal-field instability (Fig. \ref{Fig5}c). We can take advantage of the Beer-Lambert law to reconstruct the 3D shape of the pattern remembering that the thickness of the dense phase layer is proportional to the value of \(-\ln(I/I_0)\) (Fig. \ref{Fig5}d). Furthermore is possible to confirm that the profile of the perturbation is sinusoidal with the amplitude increasing while increasing the B-field strength (Fig. \ref{Fig5}e). It is important to notice that the pattern is sinusoidal around the unperturbed interface only in the direction \textbf{a}\(_1 + \)\textbf{a}\(_2\) (see Fig. \ref{Fig5}c for reference). We can determine the overall periodicity of the pattern by looking at the intensity if the FFT (See Methods and Extended Data Fig. 7 for details of the calculations). The value of the periodicity clearly depends both from the sample studied and the B-field applied (Fig. \ref{Fig5}f). In particular, each sample clearly shows a linear dependence of the periodicity with the magnetic field applied. If we further increase the magnetic field, at a certain point, the dense phase peaks make contact with the upper glass wall. This fact can be confirmed by looking at the histogram of the pixel intensity in the image: suddenly a peak in the range of the darker pixels appears corresponding to the value of the whole thickness of the capillary occupied by the dense phase (Extended Data Fig. 8b). This phenomenon sets a limit for the study of the small perturbations in the system but, at the same time, allows us to calibrate the Beer-Lambert law and extract precise values of the amplitude of the pattern (Fig. \ref{Fig5}g). The data for the amplitude calculated using eq. \ref{equation4} indicate that the three samples studied behave very similarly. 
Despite the differences between the observed pattern and the classical hexagonal pattern for the normal-field instability we should expect that the basic physical principles for the onset of the instability are the same.
Furthermore, even if eq. \ref{equation1} and \ref{equation2} were calculated for an infinite layer of ferrofluid it is important to notice that for a ferrofluid with a magnetic susceptibility \(\chi \sim 0.05\) and depth of the layer less than the pattern periodicity, the critical magnetic field and the critical wave number are altered less than 2\%. Moreover, if the depth is larger than the pattern periodicity this effect can be completely ignored. \cite{richter2009surface} This observation apply very well to our case since for all our samples \(\chi \sim 0.06\) but the thickness of the dense phase is always larger of the amplitude of the pattern implying that we don't need to take into account the effects of the layer thickness.
Therefore, starting from the periodicity values near the critical field for the three different samples we can use eq. \ref{equation1} and \ref{equation2} to extract the approximate value of the interfacial tension and the theoretical value of the critical field for each sample to check the validity of the model comparing it with the experimental one. The values of the average periodicity for the three different samples right after the onset of the instability are:  244.3 \(\pm\) 0.8 \(\upmu\)m for sample 0,	225.7 \(\pm\) 0.8 \(\upmu\)m for sample 1 and 189.4 \(\pm\) 0.7 \(\upmu\)m for sample 2.
Taking into consideration these values the calculated interfacial tensions are 0.9 \(\pm\) 0.1 \(\upmu\)N/m for sample 0, 0.6 \(\pm\) 0.1 \(\upmu\)N/m for sample 1 and 0.45 \(\pm\) 0.09 \(\upmu\)N/m for sample 3. We can use these values of interfacial tension to calculate the critical value of B-field at which we should expect the instability to take place. Using eq. \ref{equation2} the critical B-field is calculated as 5.0 \(\pm\) 0.6 mT for sample 0, 4.3 \(\pm\) 0.6 mT for sample 1 and 4.4 \(\pm\) 0.6 mT for sample 2. All the calculated values are very close to the experimental value of 4.8 mT confirming the good agreement with the model (for further details on the calculations of these values see Supplementary Note S5). \\

\begin{figure}[H]
\centering
\includegraphics{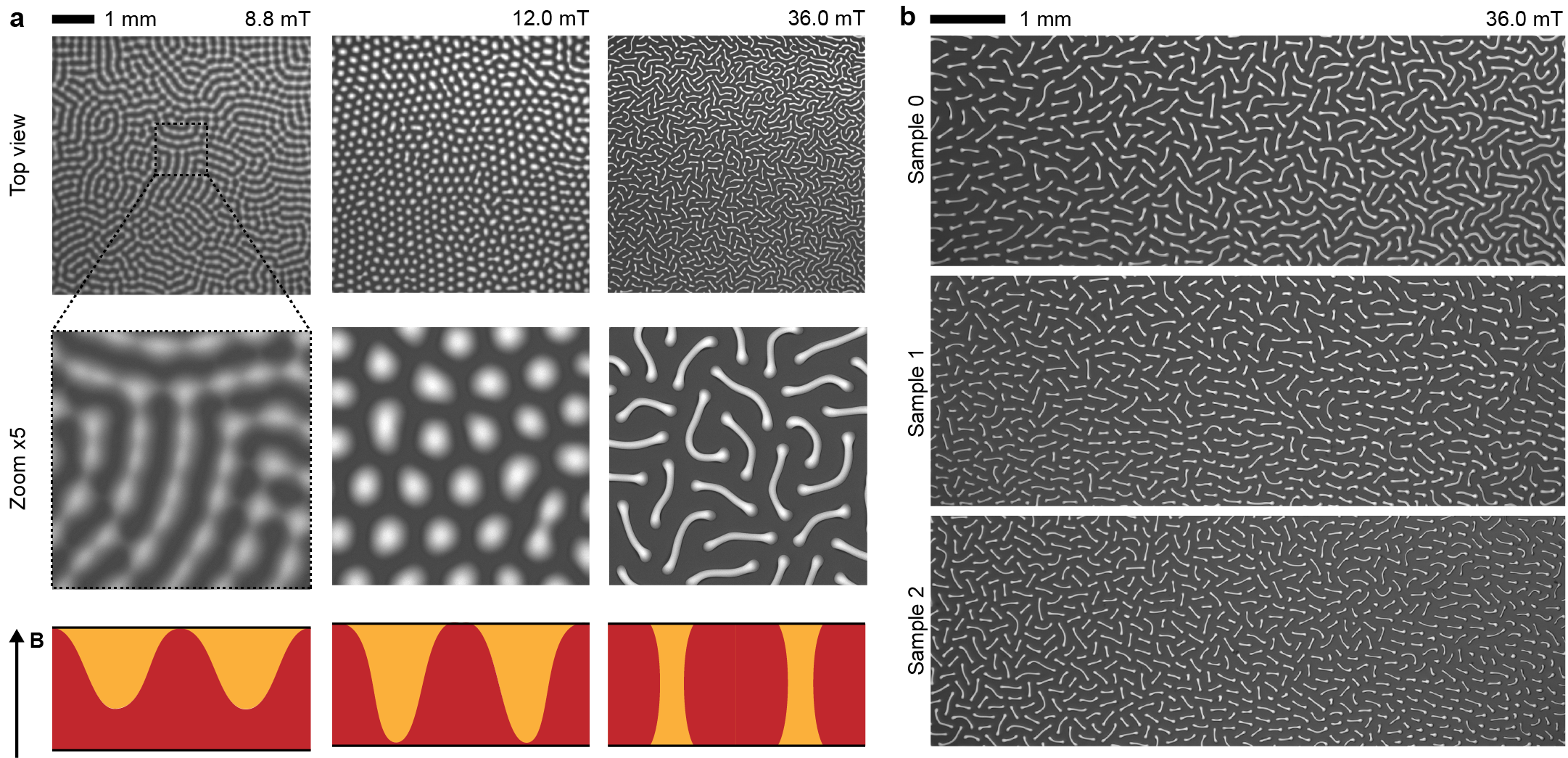}
\caption{\textbf{Transitioning from the normal-field instability pattern to the labyrinthine pattern.}
\textbf{a,} Images of the view along the z-axis in function of the magnetic field applied (first row), zoom of the central area of the images (second row) and scheme of the proposed shape of the interface between the two phases for a section along the x-z plane. 
\textbf{c,} Behavior of the new pattern at 36.0 mT in three different samples.}
\label{Fig6}
\end{figure}

Increasing the magnetic field up to a certain value leads the dense phase peaks to make contact with the upper glass wall (Fig. \ref{Fig6}a, first column). Increasing the magnetic field over this point causes the progressive division of the light phase into small wells surprisingly well dispersed in size. In this state the wells are also organized in an hexagonal pattern (Fig. \ref{Fig6}a, second column). Still increasing the field also induces the light phase to elongate enough to touch the glass wall at the other side. Having no more space to develop vertically the magnetic instability induces horizontal deformation of the light phase wells, in an effect very similar to the one of the labyrinth instability for confined magnetic fluids (Fig. \ref{Fig6}a, third column). Note that since the contact angle for the dense phase is approximately 180 \(^\circ\) when the light phase makes contact with the lower glass wall the shape became symmetric respect of the central axis of the capillary (see Methods and Extended Data Fig. 2 for details on the contact angle measure). If from this point we decrease the magnetic field step by step we can obtain the same exact states with only small differences in the overall arrangement of the pattern (see Extended Data Fig. 9 for extra images of the patterns). These novel magnetic field instability patterns link conceptually the normal-field pattern to the labyrinthine one. The two instabilities are clearly linked to the forces and pressures arising from the magnetization differences between the two phases and to the thicknesses ratio between the two materials. This suggests that the two phenomena can be described as different states of the same overall behavior. \\
In conclusion we have fully characterized the miniaturized magnetic instabilities in a new ferrofluidic system based on the so called ATPS approach. At this record scale the normal-field instability seems to loose the almost perfect long range ordering typical of the classical system in favour of a domain like local ordering. The periodicity of the magnetic patterns is shown to go down to \(\sim\)200 \(\upmu\)m which is two order of magnitude smaller than the periodicity obtained in the classical system. We have also observed novel patterns generated by the magnetic field related to the balance between confinement and magnetic field strength. Beyond the novelty of these observations, we foresee that this approach can also be used to evaluate interfacial tension values as low as \(\sim\)200 nN/m which is very near the smallest values that can be measured with other techniques.\cite{lau2020using} From another point of view this system could be the optimal case of study to test the relation between the onset of the normal-field instability and the thermal capillary waves. With so low interfacial tensions the thermal capillary waves are known to increase their amplitude up to the optical observable limit. \cite{aarts2004direct} Having the possibility to observe in the same system both thermal capillary waves and magnetic instabilities could allow to study the interplay between the two in a complete, non-ambiguous way.  

\section*{Methods}

\justify \textbf{Materials}\\
Iron (III) chloride hexahydrate ($\mathrm{FeCl_3\cdot6H_2O}$, $\geq$ 99\%, 31232, Sigma-Aldrich), iron (II) chloride tetrahydrate ($\mathrm{FeCl_2\cdot4H_2O}$, $\geq$ 99\%, 44939, Sigma-Aldrich), iron (III) nitrate nonahydrate ($\mathrm{Fe(NO_3)_3\cdot9H_2O}$, $\geq$ 98\%, 216828, Sigma-Aldrich), hydrochloric acid (HCl, 37\%, 258148, Sigma-Aldrich),  ammonium hydroxide ($\mathrm{NH_4OH}$, 28-30\% $\mathrm{NH_3}$ basis, 221228, Sigma-Aldrich), nitric acid ($\mathrm{HNO_3}$, 70\%, 438073, Sigma-Aldrich), acetone ($\geq$ 99.8\%, A/0606/17, Fisher Scientific), diethyl ether ($\mathrm{(C_2H_5)_2O}$, $\geq$ 99.8\%, A/0606/17, Sigma-Aldrich), polyethylene glycol (PEG 35000, $\geq$ 99.9\%, 81310, Sigma-Aldrich), dextran (Dextran T500, $\sim$ 95\%, 40030,  Pharmacosmos), glass capillaries (Vitrocom Inc. hollow rectangular capillaries fabricated in borosilicated glass, open ends, flame polished: 0.05 \(\times\) 1.00mm, wall thickness 0.05 mm, tolerances \(\pm\) 10\%, product ID 5015-050; 0.10 \(\times\) 2.00mm, wall thickness 0.10 mm, tolerances \(\pm\) 10\%, product ID 5012-050; 0.20 \(\times\) 4.00mm, wall thickness 0.20 mm, tolerances \(\pm\) 10\%, product ID 3524-050; 0.40 \(\times\) 8.00mm, wall thickness 0.40 mm, tolerances \(\pm\) 10\%, product ID 2548-050), UV curable adhesive (Norland Optical Adhesive 61).

\justify \textbf{Synthesis of citrated maghemite NPs dispersion in water (ferrofluid)}\\
\textit{Formation of maghemite NPs:} maghemite (\(\gamma - \textrm{Fe}_2\textrm{O}_3\)) NPs were synthetized by coprecipitation of \(\textrm{Fe}^{2+}\) and \(\textrm{Fe}^{3+}\) salts in an alkaline medium. \cite{massart1981preparation,talbot2018ph} In detail, 210 g of \(\textrm{FeCl}_3 \cdot 6\textrm{H}_2\textrm{O}\) (0.78 mol) were added to distilled water and dissolved by magnetic stirring (total volume of 1.5 L). A second solution is prepared by dissolving 90 g of \(\textrm{FeCl}_2 \cdot 4\textrm{H}_2\textrm{O}\) (0.45 mol) in an acidic solution (50 ml of HCl 37\% w/w added to 250 ml of distilled water). The two solutions are mixed under mechanical stirring (400 rpm) and 360 ml of \(\textrm{NH}_4\textrm{OH}\) (28-30\% w/w of NH\(_3\)) is poured quickly into the iron solution. The solution is stirred for 30 min and then washed twice by magnetic decantation (the supernatant is discarded). \\
\textit{Acidification of the NPs:} a solution of HNO\(_3\) (2 mol/L) is poured over the decantated sediment until 1 L is reached. Afterwards the solution is stirred at 400 rpm for 30 min and then decanted on a strong magnet (the supernatant is discarded). \\
\textit{Oxidation of the NPs:} a solution is prepared by dissolving 270 g of Fe(NO\(_3\))\(_3\) \(\cdot\) 9H\(_2\)O (0.67 mol) in 400 ml of distilled water. The solution is boiled and then poured over the NPs. Boiling is maintained for 30 min and after magnetic decantation, the supernatant is discarded. 
A solution of HNO\(_3\) (2 mol/L) is poured over the NPs until 1 L is reached and the solution is stirred at 400 rpm for 10 min. Then the solution is decanted on a strong magnet and the supernatant is discarded. The precipitate is then washed 3 times with acetone and 2 times with ether. Distilled water is added until 600 ml is reached and the remaining ether is evaporated at 40 \(^{\circ}\)C for a few hours. \\
\textit{Stabilization of NPs at neutral pH with sodium citrate:} 14 g of sodium citrate (Na\(_3\)C\(_6\)H\(_5\)O\(_7\), 4.76 \(\times\) 10\(^{-2}\) mol) are added to the ferrofluid, which is then stirred for 30 min at 80 \(^{\circ}\)C. The solution is then washed twice with acetone and twice with ether. Afterwards, the particles are dispersed in distilled water until 100 ml is reached and the remaining ether is evaporated at 40 \(^{\circ}\)C. Finally, the ferrofluid is filtered through 0.2 um pores and distilled water is added to reach a final volume of 135 mL. 
The volume fraction of NPs in this stock dispersion was determined to be approximately 4.4 \% (Supplementary Note S1).

\justify \textbf{Characterisation of the synthesised citrated ferrofluid}\\
\textit{Transmission electron microscopy (TEM):} NP morphology and size distribution were determined using a transmission electron microscope (JEOL JEM-2800, 200 kV). TEM sample was prepared by diluting the NP stock dispersion with water approximately in ratio 1:1000 and pipetting a droplet on a TEM grid with a holey carbon film (Agar Scientific S147-4) and allowing the water to evaporate. From individual TEM images (Extended Data Fig. 1a), the NP size distribution (Extended Data Fig. 1b) was determined using ImageJ software measuring the diameter of each particle manually in a random direction (Approximately 430 NPs were measured). The data relative to the diameter of the NPs were interpolated with a log-normal distribution of equation:
\begin{equation}
    y = y_0 + \frac{A}{x S \sqrt{2\pi}} \exp \left(-\frac{(\ln(x) - M)^2}{2 S^2}\right)
\end{equation}
giving the results reported in the table: 

\begin{table}[h!]
    \centering
    \caption{Results of the interpolation of the NP size data with the log-normal distribution of eq. 2}
\begin{tabular}{|c|c|c|c|}
\hline
\(y_0\)& \(A\) & M & S\\
\hline
0 \(\pm\) 2   & 548 \(\pm\) 35 & 2.13 \(\pm\) 0.01 & 0.33 \(\pm\) 0.02 \\
\hline
\end{tabular}
\end{table}
\noindent
From these results we can calculate the mean (\(\mu\)) and standard deviation (\(\sigma\)) as follows:
\begin{equation}
    \mu = e^{M+S^2/2} , \qquad \sigma^2 = e^{S^2+2M} \left( e^{S^2} -1 \right) 
\end{equation}
The results of the calculations are: \(\mu = 8.9 \pm 0.1 \) nm and \(\sigma = 3.00 \pm 0.09 \) nm

\justify \textit{Magnetometry:} Magnetic properties of the citrated ferrofluid were measured with a vibrating sample magnetometer (QuantumDesign PPMS VSM). A 3 cm long capillary tube with 0.22 mm inner diameter was filled with ca. 1 ${\mathrm{\upmu l}}$ of the sample to be measured. The filled capillary was weighed to measure the weight of the citrated ferrofluid and a photo was taken with a size reference for exact volume determination. The capillary was sealed with UV curable adhesive (Norland Optical Adhesive 61) and cured under UV lamp (Thorlabs Solis 365C). The paramagnetic background from sample holder was evaluated by fitting the data at highest field and then subtracted from the measured data  (Extended Data Fig. 1c). Average NP magnetic moment, NP number and saturation magnetization were obtained by fitting the measured magnetization loop with Langevin theory for superparamagnetism (see Supplementary Note S3). The magnetic susceptibility was obtained by linear interpolation of the data at low magnetic field.

\justify \textit{Density:}  The density of the initial citrated ferrofluid was determined. 100 uL of solution were pipetted with an Eppendorf Multipette E3x equipped with Combitips advanced 0.1 ml and weighed with an Ohaus Pioneer analytical balance. Each measurement was performed at room temperature (23 \(\pm\) 1 \(^{\circ}\)C) in triplicate. The density of distilled water was verified beforehand to validate the obtained values (density measured: 0.997 \(\pm\) 0.008 g/ml). The final averaged value of the density of the citrated ferrofluid was calculated to be: 1.28 \(\pm\) 0.08 g/ml. 

\justify \textbf{Preparation and characterization of the FF-ATPS samples}
\justify \textit{Sample preparation:} The main FF-ATPS sample (Sample 0) is obtained by mixing 4 ml of the synthesized citrated ferrofluid with 36 ml of a concentrated PEG-Dextran solution (PEG: 1.8000 \(\pm\) 0.0002 mass\% ; Dextran: 3.9284 \(\pm\) 0.0002 mass\%, more details on the exact quantities mixed in Supplementary Note S2).
The concentrated PEG-Dextran solution was obtained by weighting the wanted amounts of PEG, Dextran and water in a centrifuge tube and mixing it at 300 rpm overnight. The citrated ferrofluid was then mixed and the whole dispersion was put in a vortex mixer at 300 rpm for 10 min and then centrifuged (Beckman Coulter / Allegra X-22R) for 2 hours at 5000 g. 
The other samples (Samples 1, 2, 3 and 8) were obtained as follows: four samples of 2 ml of the main batch were extracted after vigorous mixing for 2 minutes to to be sure that each batch would have the same light/dense phase ratio, then different amounts of water were added to each of the sample. The water dilution for each sample is the following:

\begin{table}[h!]
    \centering
\caption{water added to prepare each FF-ATPS dilution.}
\begin{tabular}{|c|c|c|c|c|}
\hline
Sample & 1 & 2 & 3 & 8 \\
\hline
Water dilution (vol\%) & 0.927 \(\pm\) 0.003 & 1.833 \(\pm\) 0.003 & 2.756 \(\pm\) 0.003 & 7.035 \(\pm\) 0.009  \\
\hline
\end{tabular}
\end{table}
\noindent
All mass measurements during the preparation were carried out after discharging the object to be weighted by passing it through an electrostatic gate.

\justify \textit{Density:} the density of both the light and the dense phase was measured for samples 0, 1, 2, 3 and of the single phase of sample 8. The measurements were carried out following the same procedure described for the citrated ferrofluid. The light and dense phase of samples 0, 1, 2 and 3 were separated carefully pipetting the top and the bottom of each sample batch. The results of the density for each sample are summarized in the table below.
\begin{table}[h!]
    \centering
\caption{Measured density for each FF-ATPS dilution.}
\begin{tabular}{|c|c|c|c|c|c|}
\hline
Sample & 0 & 1 & 2 & 3 & 8 \\
\hline
Light phase density (g/ml) & 1.01 \(\pm\) 0.01 & 1.01 \(\pm\) 0.01 & 1.01 \(\pm\) 0.01 & 1.02 \(\pm\) 0.01 & \multirow{2}{*}{1.05 \(\pm\) 0.01 }  \\
\cline{1-5}
Dense phase density (g/ml) & 1.07 \(\pm\) 0.01 & 1.07 \(\pm\) 0.01 & 1.07 \(\pm\) 0.01 & 1.06 \(\pm\) 0.01 & \\
\hline
\end{tabular}
\end{table}
\noindent
Note: the density of sample 8 is closer to the values of the dense phases because the light/dense phase volume ratio in each sample is 1:3.

\justify \textit{Magnetometry:} the magnetic properties of both the light and the dense phase were measured for samples 0, 1, 2, 3 and of the single phase of sample 8. The measurements were carried out following the same procedure described for the citrated ferrofluid. Average NP magnetic moment, NP number and saturation magnetization were obtained by fitting the measured magnetization loops with Langevin theory for superparamagnetism and the magnetic susceptibility was obtained by linear interpolation of the data at low magnetic field. (see Extended Data Fig. 2b,c and Supplementary Note S3 for further details).

\justify \textit{Interfacial tension and contact angle:} 
The interfacial tension between the PEG and DEX rich phase of the FF-ATPS samples is approximated using a sessile drop method. \cite{atefi2014ultralow} \\
\textit{Formation of sessile drops.} A quartz cuvette with a thickness of 1 mm is filled with the light phase of a FF-ATPS sample. Then, a small drop of dense phase is added. The volume of the dense rich phase should be large enough to get a puddle shape and not a spherical shape (see Extended Data Fig. 2) but not too large to be able to observe the whole drop with the objective. For samples 0 to 3, the volume of the drops is 2, 1, 0.5, and 0.5 \(\upmu\)l, respectively. The cuvette is placed horizontally on the microscope stage. Then, the drop is moved with a small magnet so that it is suspended on the upper part of the cuvette. The magnet is removed and the drop falls down on the lower part of the cuvette. \\
\textit{Setup and image acquisition.} Spreading drops on the lower part of the cuvette are visualized with an inverted microscope (Nikon Eclipse Ti) equipped with a 2x objective. Pictures of the puddle shaped drop are taken using a SCMOS camera (Andor Zyla). \\
\textit{Data analysis.} The interfacial tension \(\gamma\) can be determined using:
\begin{equation}
    \gamma = \frac{\Delta \rho g e_c^2}{4 \sin^2 (\theta_E/2) }
\end{equation}
where \(\Delta \rho\) is the density difference between the two phases, \(g\) is the gravitational acceleration, \(e_c\) is the thickness of the puddle and \(\theta_E\) is the contact angle (see Extended Data Fig. 2d). Because of the colored ferrofluid it is difficult to determine the contact angle. However, in the case of ATPS samples the contact angle is close to 180\(^\circ\) (see Extended Data Fig.2e). By approximating \(\theta_E \sim\)180\(^\circ\) (the contact angle is slightly overestimated), the previous equation becomes:
\begin{equation}
      \gamma = \frac{\Delta \rho g e_c^2}{4}
\end{equation}
where \(e_c =\) volume of the drop / area of the drop. Measuring the area of the droplet we can approximate \(\gamma\). While the drop spreads, it also dewets, which creates holes in the puddle. The area of the puddle is measured when the whole drop falls down. It means that the value for the area is slightly overestimated and \(e_c\) is minimized. The results of the determinations of \(\gamma\) for the four samples are:

\begin{table}[h!]
    \centering
    \caption{Results of the approximative interfacial tension determination for the four samples.}
\begin{tabular}{|c|c|c|c|c|}
\hline
Sample &  0 &  1 &  2 &  3\\
\hline
Interfacial tension (\(\upmu\)N/m) & \(\sim\)4   & \(\sim\)0.2 & \(\sim\)0.1 & \(\sim\)0.1 \\
\hline
\end{tabular}
\end{table}

\justify \textit{Light absorption:} the light absorption of both the light and the dense phase was measured for samples 0, 1, 2, 3 and of the single phase of sample 8. The measurement of the light intensity was performed averaging the pixel values in an area of 1.00\(\times\) 1.00 mm with center at 2.00 mm distance from the interface in a 0.10\(\times\)2.00 mm capillary tube. The single phase sample (8) intensity value was measured in the middle of the capillary always averaging the pixel values in an area of 1.00\(\times\) 1.00 mm. This last intensity value was used as reference for the concentration in the Beer-Lambert model. This choice even if not exact since it does not take into account the extra dilution of the system allows us to give an approximate comparison of the partition of the NPs in the two phases for the other samples.

\justify \textit{Capillary filling and sealing:} before filling the capillaries the samples were vigorously mixed for 2 min. After that a certain amount of the samples was extracted depending of the thickness of the capillary to fill (2 \(\upmu\)l for the 0.05\(\times\) 1.00 mm, 8 \(\upmu\)l for the 0.10\(\times\) 2.00 mm, 20 \(\upmu\)l for the 0.20\(\times\) 4.00 mm and 80 \(\upmu\)l for the 0.40\(\times\) 8.00 mm). After filling, the capillary was sealed with UV curable adhesive (Norland Optical Adhesive 61) and cured under UV lamp (Thorlabs Solis 365C). Each filled and sealed capillary was left to rest vertically overnight to allow phase separation before using them for the measurements.

\justify \textbf{Experimental setup for microscopic observation of the FF-ATPS in the glass capillaries under magnetic field.}
\justify \textit{Magnetic field:} the magnetic field was generated and controlled as before.\cite{cherian2020Electroferrofluids} In Brief a pair of small electromagnetic coils (GMW 11801523 and 11801524) connected to DC power supply (BK Precision 9205 Multi-Range DC Power Supply) was used to generate uniform magnetic fields. The magnetic field between the coils was calibrated using a 3-axis teslameter (Senis 3MTS).
\justify \textit{Microscopy:} The capillaries were illuminated in transmitted light configuration using an LED light source (Thorlabs MCWHLP1), collimator (Thorlabs COP4-A Zeiss) and a light diffuser. All images were captured using a 4x finite-conjugate objective lens (Nikon 4x/0.25 160/- WD25) or a 1x Telecentric Gauging Lens (Melles Griot Macro Invaritar 1x 59LGM601) connected to a 5 MP grayscale camera (Basler acA2440-75um). Image length scale was calibrated using a calibration target (Thorlabs R1L3S2P).

\justify \textbf{Methods for the meniscus elongation study.}
\justify \textit{Samples, capillaries preparation  and magnetic field protocol:} all the meniscus elongation measurements were performed by looking at the FF-ATPS sample 0 in a rectangular glass capillary of 0.20\(\times\) 4.00 mm section. The capillary was positioned vertically, parallel to the gravitational acceleration. The magnetic field was applied parallel to the gravitational acceleration, with progressive steps of 0.8 mT each from 0.0 mT to 8.0 mT. At each step, 2 min were waited from the increase of the magnetic field to the image taking to guarantee that the image was relative to an equilibrium position of the meniscus. 
\justify \textit{Light intensity profiles collection and analysis:} the intensity along the \textit{z} direction \(I(z)\) was measured from the image by averaging along the x direction (0.01 mm large average area) with the Profile Plot command of ImageJ/Fiji software. \cite{rueden2017imagej2, schindelin2012fiji} The area was selected in the same region of the capillary for all the measurements. The data were then exported in OriginPro 2020b for the analysis. As a reference for the thickness of the meniscus a constant fit of the data in the dense phase region was performed (Extended Data Fig. 3). Following the Beer-Lambert law approach all the data were divided by the intensity measured in the dense phase region and then the logarithm of the intensity ratio (\(-\ln(I/I_0)\)) was considered (see eq. \ref{equation4}). Once calibrated the same formula knowing the thickness of the capillary was possible to translate the intensity values into profile sections in the \textit{x} direction (Extended Data Fig. 3d,e ). The data very near to the interface were deleted since the Beer-Lambert law cannot be applied in regions were the light is focused or scattered by the interface itself (see ``refraction region'' in Extended Data Fig. 3b). A parabolic fitting was then performed to complete the shape of the profile were the data nearest the top of the meniscus were deleted. Among different possibilities for the area in the \textit{z} direction to be selected to perform the parabolic fit we choose a 0.005 mm wide region since in this way the reconstruction did not enter in the light phase region.

\justify \textbf{Methods for 2.5D instability study.}
\justify \textit{Samples, capillaries preparation and magnetic field protocol:} the measurements were performed changing both samples and capillaries. Samples 0, 1, 2, 3 and 0.05\(\times\) 1.00 mm, 0.10\(\times\) 2.00 mm, 0.20\(\times\) 4.00 mm, 0.40\(\times\) 8.00 mm capillaries were used. The capillaries were positioned vertically, parallel to the gravitational acceleration. The magnetic field was applied parallel to the gravitational acceleration with progressive steps of different intensity. At each step, 2 min were waited from the increase of the magnetic field to the image taking to guarantee that the image was relative to an equilibrium position of the instability pattern. 
The magnetic field is applied in a cycle starting from 0 mT then increasing the field up to a maximum value of 36.0 mT and then decreasing the field to 0 mT again. 

\justify \textit{Pattern analysis:} images of the instability pattern were analyzed with a custom MATLAB R2020a software. For each image the software located the profile of the interface with the \textit{edge} function and then converted in \textit{x-y} position data. This profile was then interpolated with a Fourier series up to the fifth order. From the results of the interpolation was possible to extract both the periodicity value and the amplitude of the pattern by summing the amplitudes of the Fourier components (see Extended Data Fig. 4a for details on the software steps). The data near the critical point were fitted with a linear function to determine the exact value of the critical field looking at the intercept between the curves and the \(x\) axis (see Extended Data Fig. 4b, c). The results of this determination are the following:

\begin{table}[h!]
    \centering
\caption{Results of the calculations of the critical field of the changing sample data from the fit results (see Extended Data Fig. 4b).}
\begin{tabular}{|c|c|c|c|}
\hline
Sample &  0 &  1 &  2 \\
\hline
Critical B-field (mT) & 7.6 \(\pm\) 0.9  & 6.0 \(\pm\) 0.9 & 5.3 \(\pm\) 0.9 \\
\hline
\end{tabular}
\end{table}

\begin{table}[h!]
    \centering
\caption{Results of the calculations of the critical field of the changing capillary thickness data from the fit results (see Extended Data Fig. 4c).}
\begin{tabular}{|c|c|c|c|}
\hline
Thickness &  50 \(\upmu\)m &  100 \(\upmu\)m &  200 \(\upmu\)m \\
\hline
Critical B-field (mT) & 7.0 \(\pm\) 0.9  & 7.6 \(\pm\) 0.9 & 8 \(\pm\) 1 \\
\hline
\end{tabular}
\end{table}

\justify \textbf{Methods for 3D instability study.} 
\justify \textit{Samples, capillaries preparation and magnetic field protocol:} the measurements were performed with samples 0, 1 and2 in 0.40\(\times\) 8.00 mm capillaries. The capillaries were positioned horizontally, perpendicular to the gravitational acceleration. The magnetic field was applied parallel to the gravitational acceleration with progressive steps of different intensity. At each step, 2 min were waited from the increase of the magnetic field to the image taking to guarantee that the image was relative to an equilibrium position of the instability pattern. 

\justify \textit{Periodicity of the pattern analysis:} FFT of each image was performed with ImageJ/Fiji software command. \cite{rueden2017imagej2, schindelin2012fiji} Intensity Profile Plot of the FFT image was perfomed always with ImageJ/Fiji software. The data were then exported in OriginPro 2020b for the analysis. The data were then elaborated by performing a baseline fitting and subtraction in the Fourier space and then the peak of the resulting data was interpolated with a Gaussian distribution in the real space (Extended Data Fig. 7a). To calculate the correct value of periodicity from the peak position in the FFT we should multiply it by \(\sqrt{2}\) making the assumption that the FFT peaks along the \textbf{a}\(_1\) and \textbf{a}\(_2\) directions. This can be confirmed by looking at the FFT in a region with uniform pattern (Extended Data Fig. 7b, c). For an improve statistic we consider the result of the FFT on the entire image. In this case the different directions of the domains in the pattern average making the Peak of the FFT symmetric for full rotations (see Extended Data Fig. 7a).

\justify \textit{Amplitude of the pattern analysis:} each image describing the pattern at a certain magnetic field was first divided by a reference image of the sample at 0.0 mT (Image Calculator command in ImageJ/Fiji). At this point a smoothing of the image was performed by replacing each pixel with the average of its 3\(\times\)3 neighborhood (Smooth command in ImageJ/Fiji). The smoothing was done to reduce the effect of the noise in the image due to the oscillation of the pixel values of the camera. After this, maximum and minimum position of each image were located (Find Maxima command in ImageJ/Fiji) and their intensity values were measured (Measure command in ImageJ/Fiji). The data were then exported in OriginPro 2020b for the analysis. The data of minimum/maximum intensity were analyzed through a frequency histogram distribution and then the maximum value of the histogram was selected as average value of the intensity in the peaks/valleys (Extended Data Fig. 8a). These intensity values were converted into amplitude values with the Beer-Lambert law using as a reference the intensity value corresponding to the dense phase touching the upper glass wall. This value appeared in the histogram of the intensity of each image as a defined peak in the darker values after a certain magnetic field (Extended Data Fig. 8b). The last element to determine the shape of the pattern in the capillary was the position of the interface at 0 mT i.e. the thickness of the two phases in the unperturbed state. To determine this we have evaluated the volume of each phase in the capillaries by direct imaging the same capillaries in vertical position (Extended Data Fig. 8c). The results of this determination are summarized in table \ref{table7}:

\begin{table}[h!]
    \centering
\caption{Thickness at 0 mT for each of the two phases in the three samples studied (see Extended Data Fig. 8c.}
\begin{tabular}{|c|c|c|c|}
\hline
Sample & 0 & 1 & 2 \\
\hline
Light phase thickness (\(\upmu\)m) & 117 \(\pm\) 1 & 113 \(\pm\) 1 & 96 \(\pm\) 1 \\
\hline
Dense phase thickness (\(\upmu\)m) & 283 \(\pm\) 1 & 287 \(\pm\) 1 & 304 \(\pm\) 1 \\
\hline
\end{tabular}
\label{table7}
\end{table}

\clearpage

\addcontentsline{toc}{section}{\numberline{}References}

\bibliographystyle{naturemag}
\bibliography{references}


\begin{flushleft}
\justify

\medskip
\justify
\textbf{Acknowledgements}
\justify
JVIT acknowledges funding from ERC (803937) and Academy of Finland (316219). Fereshteh Sohrabi is acknowledged for assistance with VSM and TEM measurements. We acknowledge the facilities and technical support by Aalto University at OtaNano - Nanomicroscopy Centre (Aalto-NMC).

\medskip
\justify
\textbf{Author Contributions}
\justify
GB and CR synthesized and characterized the citrated ferrofluid. GB carried out the density and interfacial tension measurements. CR prepared the FF-ATPS samples, designed and constructed the magnetic field setups, carried out all measurements of the FF-ATPS samples in magnetic fields, analyzed the data, wrote the manuscript and compiled the figures. JVIT conceived the concept, performed preliminary experiments with GB and BH, and guided and supervised the experimental work, data analysis and writing of the manuscript.

\bigskip

\justify
\textbf{Additional information}
\justify
\textbf{Correspondence and requests} 
\justify
All correspondence and requests for materials should be addressed to *jaakko.timonen@aalto.fi or *carlo.rigoni@aalto.fi.

\end{flushleft}

\newpage
\input{2_ExtendedData.tex}
\newpage
\input{3_SupplementaryNotes.tex}




\end{document}

%% file: 2_ExtendedData.tex
{\large\textbf{Extended Data}}

\renewcommand{\figurename}{Extended Data Figure}
\renewcommand{\thefigure}{\arabic{figure}}
\setcounter{figure}{0}

\begin{figure}[H]
\centering
\includegraphics[width=1\textwidth]{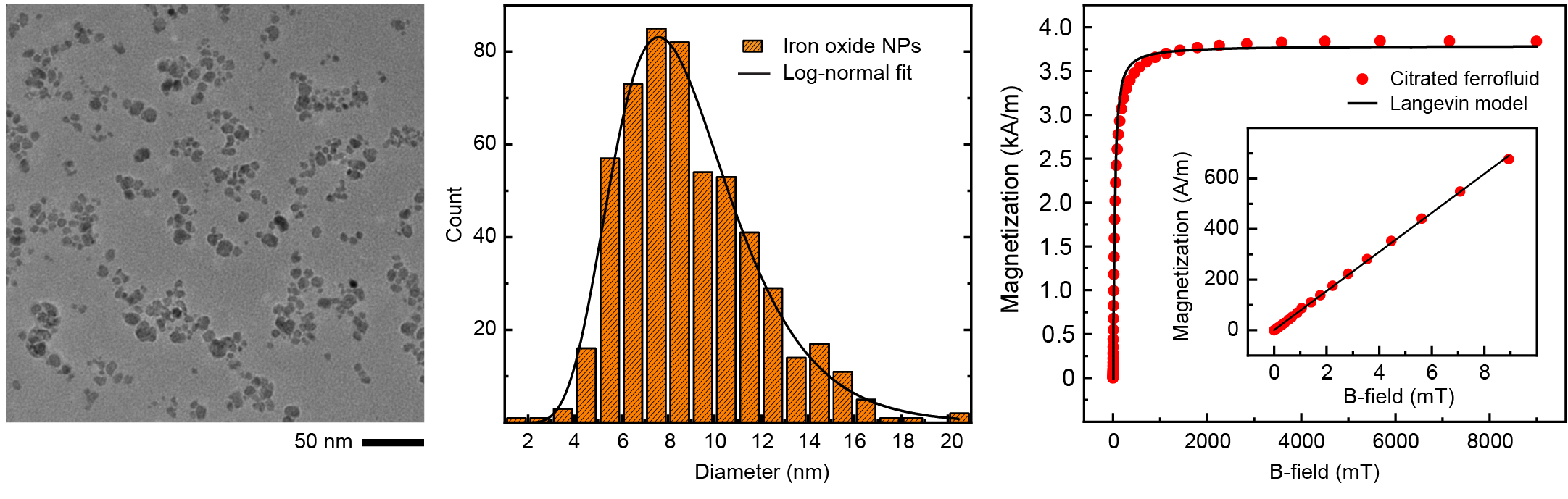}
\label{Extended_Data_Figure.01}
\caption {\textbf{Characterization of superparamagnetic NPs and the citrated ferrofluid.} \textbf{a,} A typical TEM image of the maghemite NPs stabilized with sodium citrate. 
\textbf{b,} NP diameter histogram obtained from TEM images with log-normal fit.
\textbf{c,} Magnetization curve of the stock dispersion of maghemite NPs in water and the best fit of Langevin model (See Supplementary Note S3 for model description and extracted parameters).
} 
\end{figure}

\begin{figure}[H]
\centering
\includegraphics[width=1\textwidth]{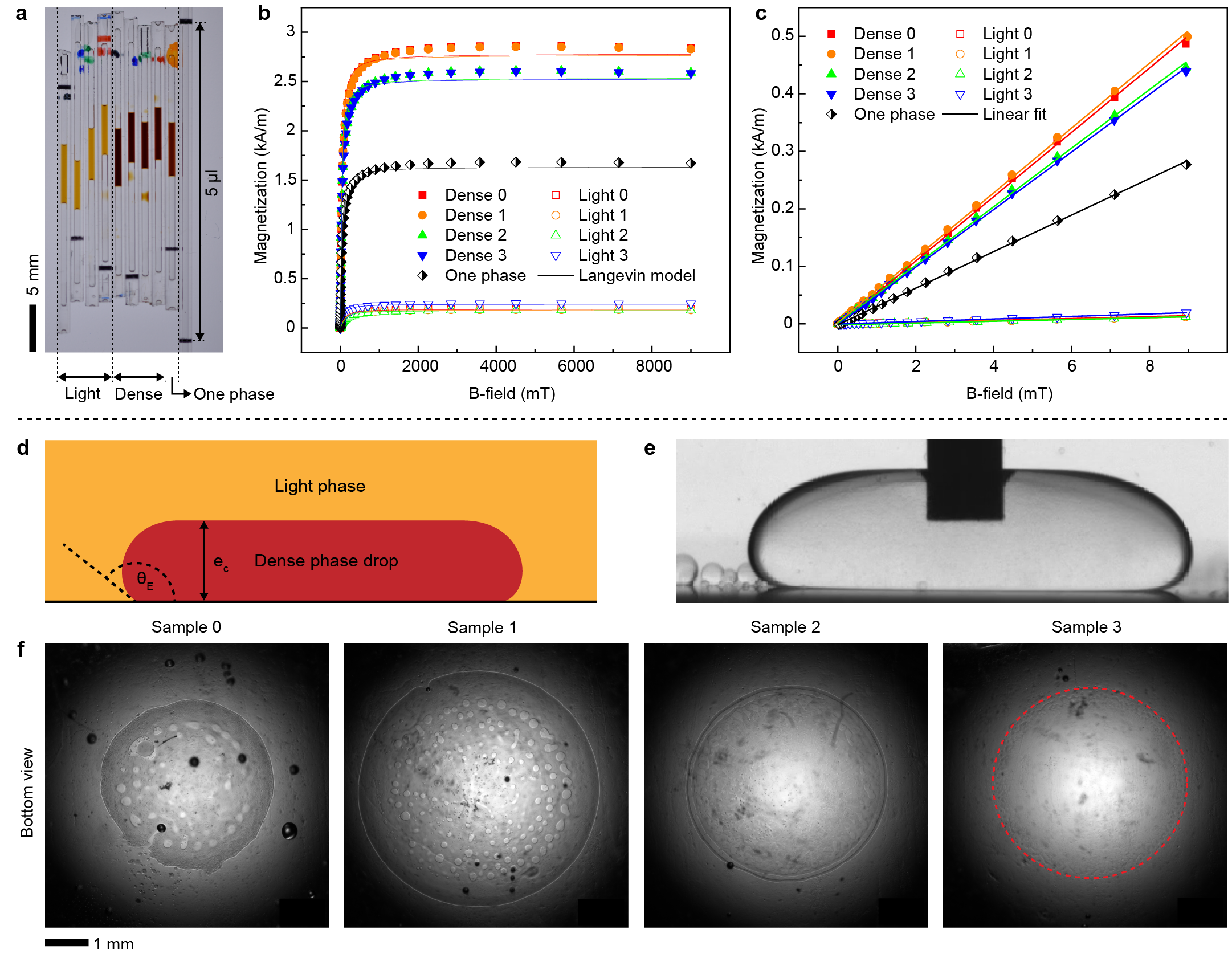}
\label{Extended_Data_Figure.02}
\caption {\textbf{Characterization of the prepared FF-ATPS samples.} \textbf{a,} Image of the samples ready for the VSM measurement. The far left capillary is used as a reference to determine the volume inside the other capillaries (the two black marks correspond to a volume of 5 \(\upmu\)l. \textbf{b,} Magnetization curves of each FF-ATPS sample prepared and the best fit of Langevin model (See Supplementary Note S3 for model description and extracted parameters). \textbf{c,} Zoom near the origin of the magnetization curves and linear fit of the data to determine the value of the magnetic susceptibility. \textbf{d,} Scheme pf a droplet in the puddle shape. \textbf{e,} Side image of a droplet of dextran-rich phase inside a PEG-rich phase in an ATPS. The image demonstrates that the contact angle of the dextran phase is near 180 \(^\circ\). \textbf{f,} Top view images of puddle droplets of dense phase in light phase for samples 0,1,2,3. The dashed lines corresponds to the area of the puddle. } 
\end{figure}

\begin{figure}[H]
\centering
\includegraphics[width=1\textwidth]{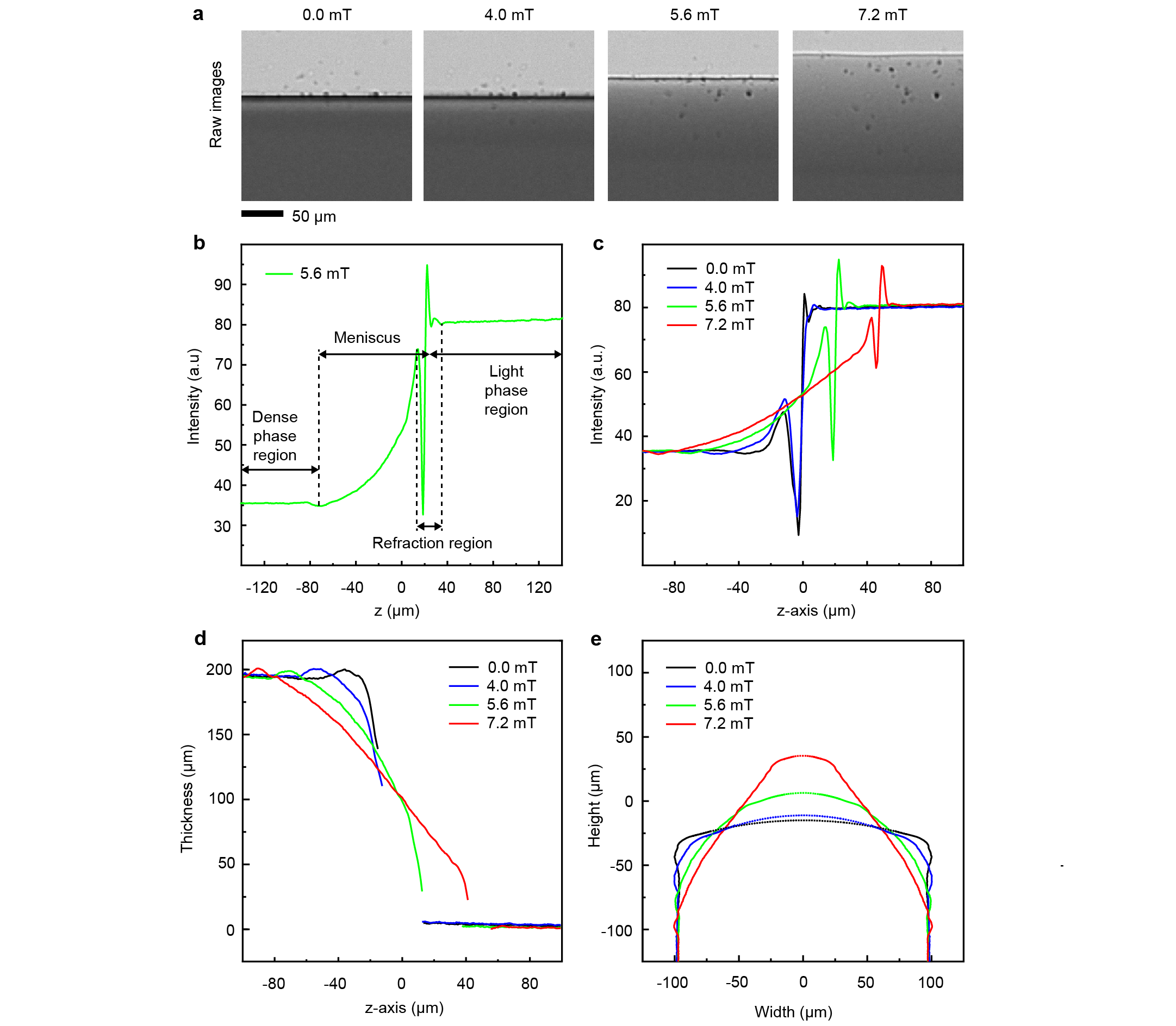}
\label{Extended_Data_Figure.03}
\caption {\textbf{Experimental methods for the meniscus elongation analysis.} \textbf{a,} Images of the meniscus elongation in sunction of the magnetic field applied. \textbf{b,} Data of the intensity profile along the vertical direction with arrows representing the regions that can be define for each profile. \textbf{c,} Data of the intensity profile along the vertical direction for four different magnetic fields. \textbf{d,} Thickness of the dense phase in function of the vertical direction. \textbf{e,} Reconstruction of the section of the interface based on the data of the thickness of the dense phase. } 
\end{figure}

\begin{figure}[H]
\centering
\includegraphics[width=1\textwidth]{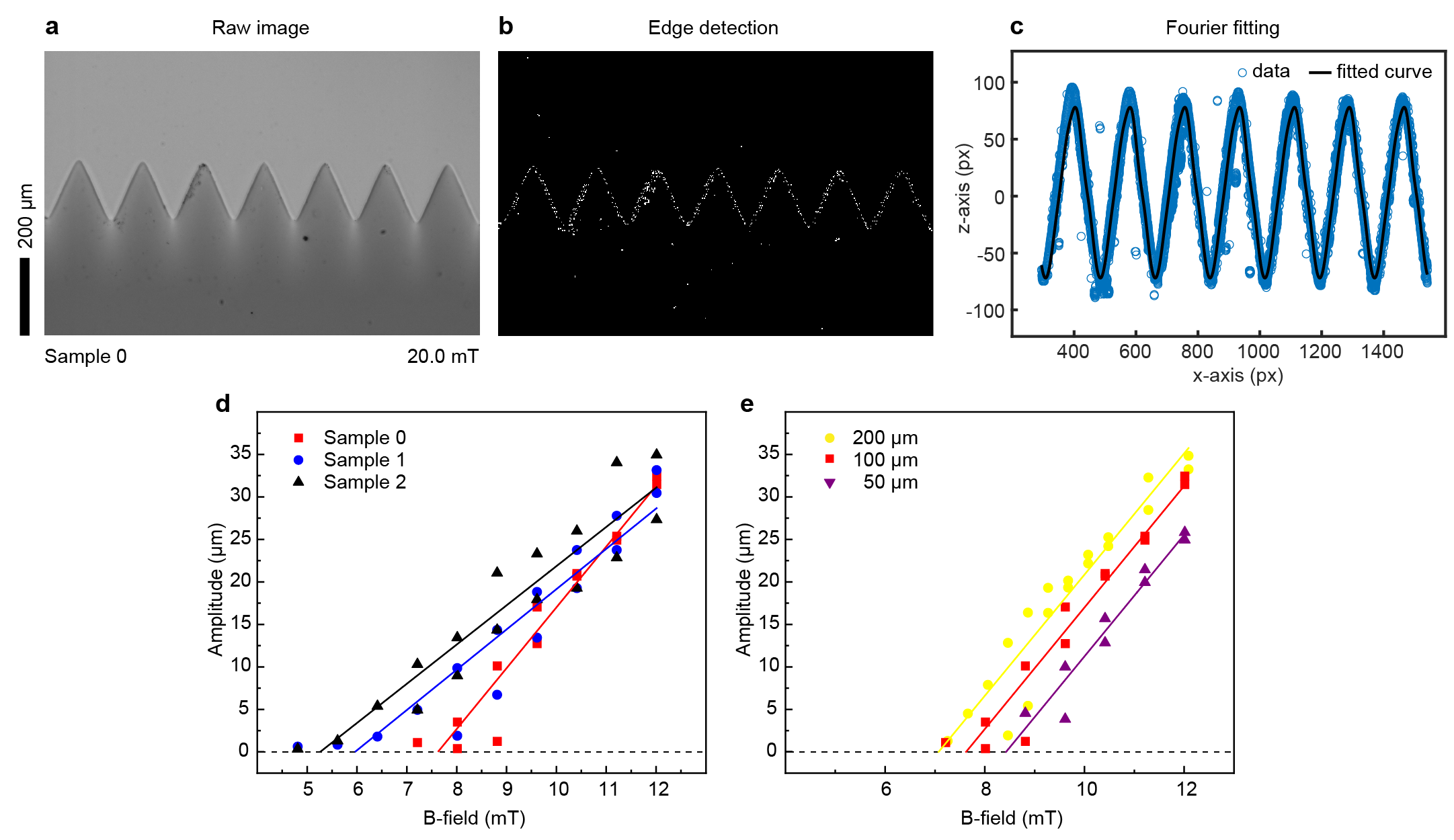}
\label{Extended_Data_Figure.04}
\caption {\textbf{Experimental methods for the meniscus magnetic instability.} \textbf{a,} Raw image of the normal field instability for sample 0 at 20.0 mT. \textbf{b,} Image of the result of the edge detection function of MATLAB  incorporated in the custom made analysis software. \textbf{c,} Data relative to the edge detection and fitting with a fourier series function up to the fifth power. } 
\end{figure}

\begin{figure}[H]
\centering
\includegraphics[width=1\textwidth]{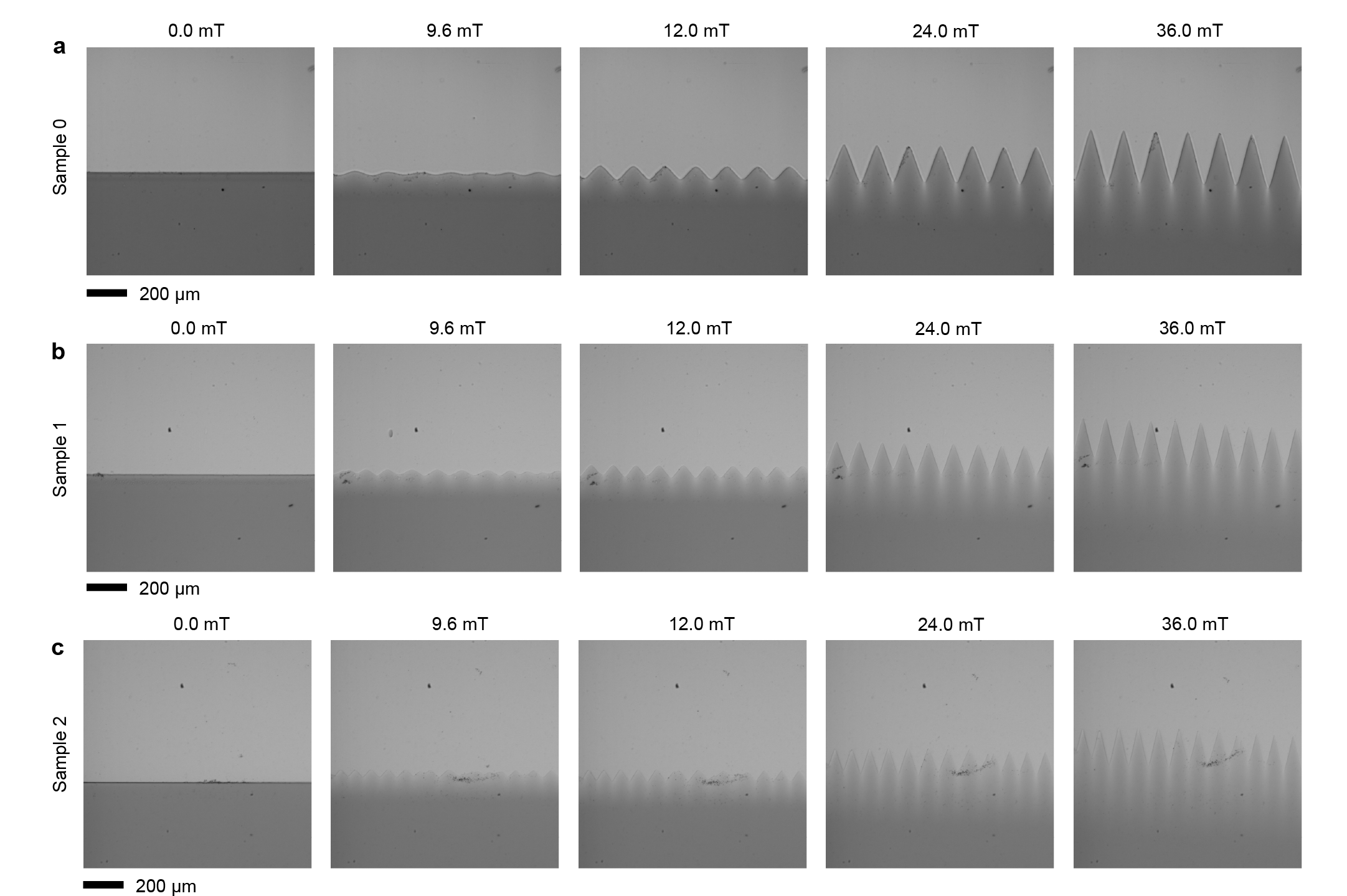}
\label{Extended_Data_Figure.05}
\caption {\textbf{Extended data for the meniscus magnetic instability: changing sample.} \textbf{a-c} Microscopy images of the normal field instability in the vertical capillaries in function of the magnetic field for three different FF-ATPS samples. } 
\end{figure}

\begin{figure}[H]
\centering
\includegraphics[width=1\textwidth]{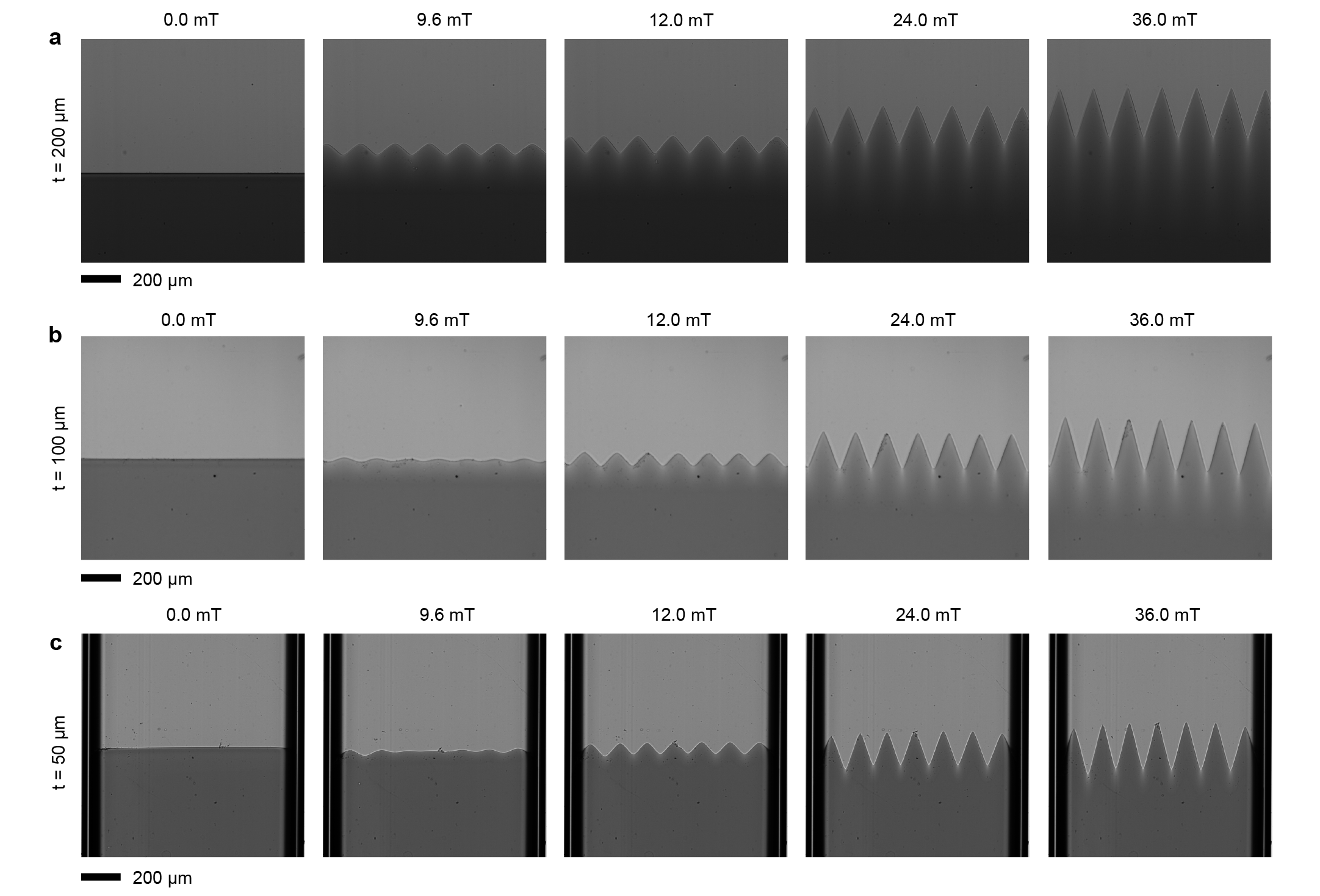}
\label{Extended_Data_Figure.06}
\caption {\textbf{Extended data for the meniscus magnetic instability: changing thickness of the capillary.} \textbf{a-c} Microscopy images of the normal field instability in the vertical capillaries in function of the magnetic field for three different capillaries thicknesses.}
\end{figure}

\begin{figure}[H]
\centering
\includegraphics[width=1\textwidth]{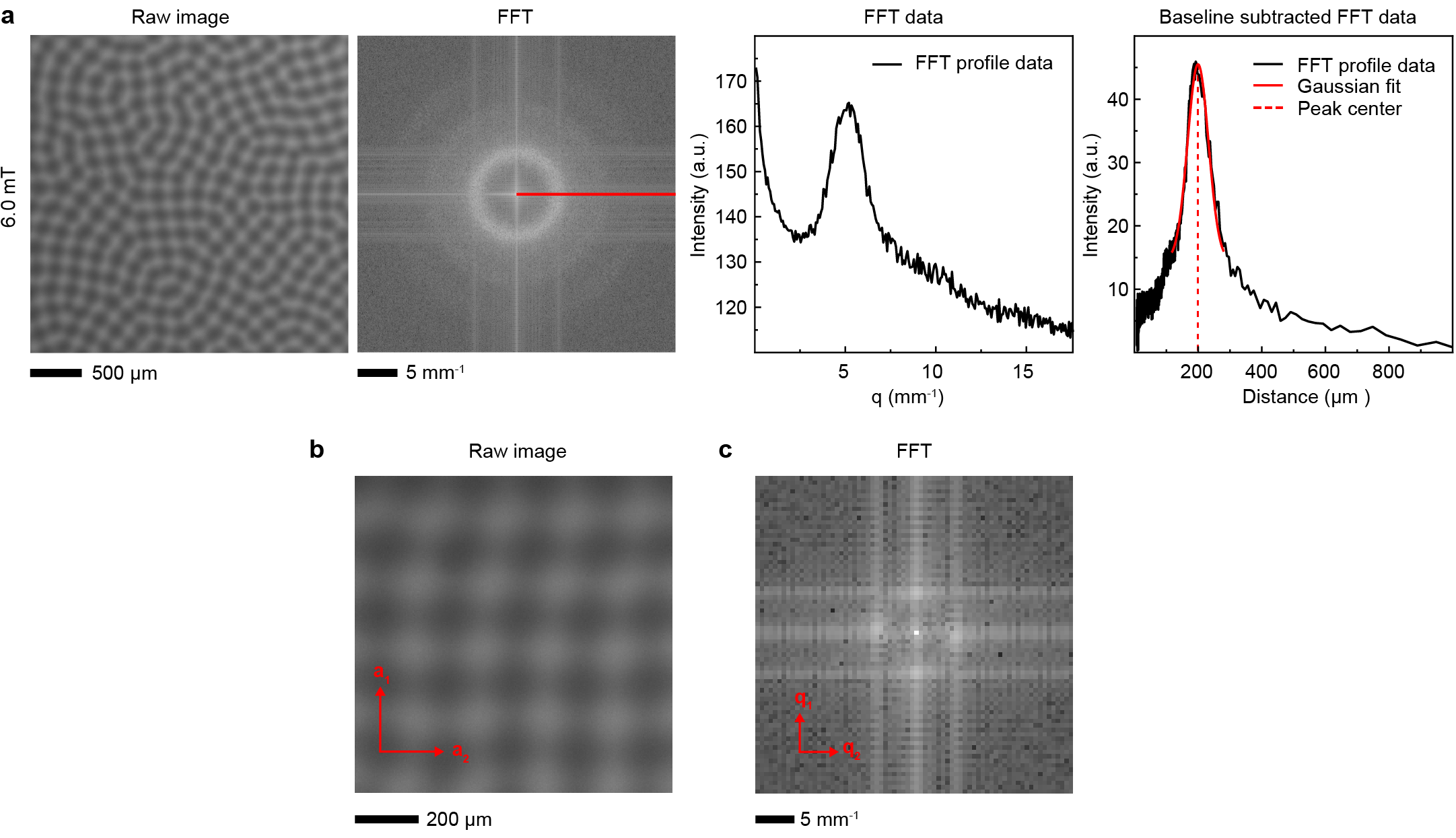}
\label{Extended_Data_Figure.07}
\caption {\textbf{Experimental methods for the periodicity analysis of the instability in the horizontal configuration.} \textbf{a,} From left to right the different steps of the analysis for the periodicity of the instability in the horizontal capillary: raw image, FFT image, intensity profile data along the direction of the red line in the FFT image, baseline subtracted data and Gaussian fitting to determine the position of the peak. \textbf{b,} Raw image of a selected area in the pattern with high regularity of the pattern. The vectors \textbf{a\(1\)} and \textbf{a\(2\)} represent the two main directions in the unit cell of the pattern lattice. \textbf{c,} FFT image of a selected area in the pattern with high regularity of the pattern (\textbf{b}). The vectors \textbf{q\(1\)} and \textbf{q\(2\)} correspond to the vectors \textbf{a\(1\)} and \textbf{a\(2\)} in the Fourier space.}
\end{figure}

\begin{figure}[H]
\centering
\includegraphics[width=1\textwidth]{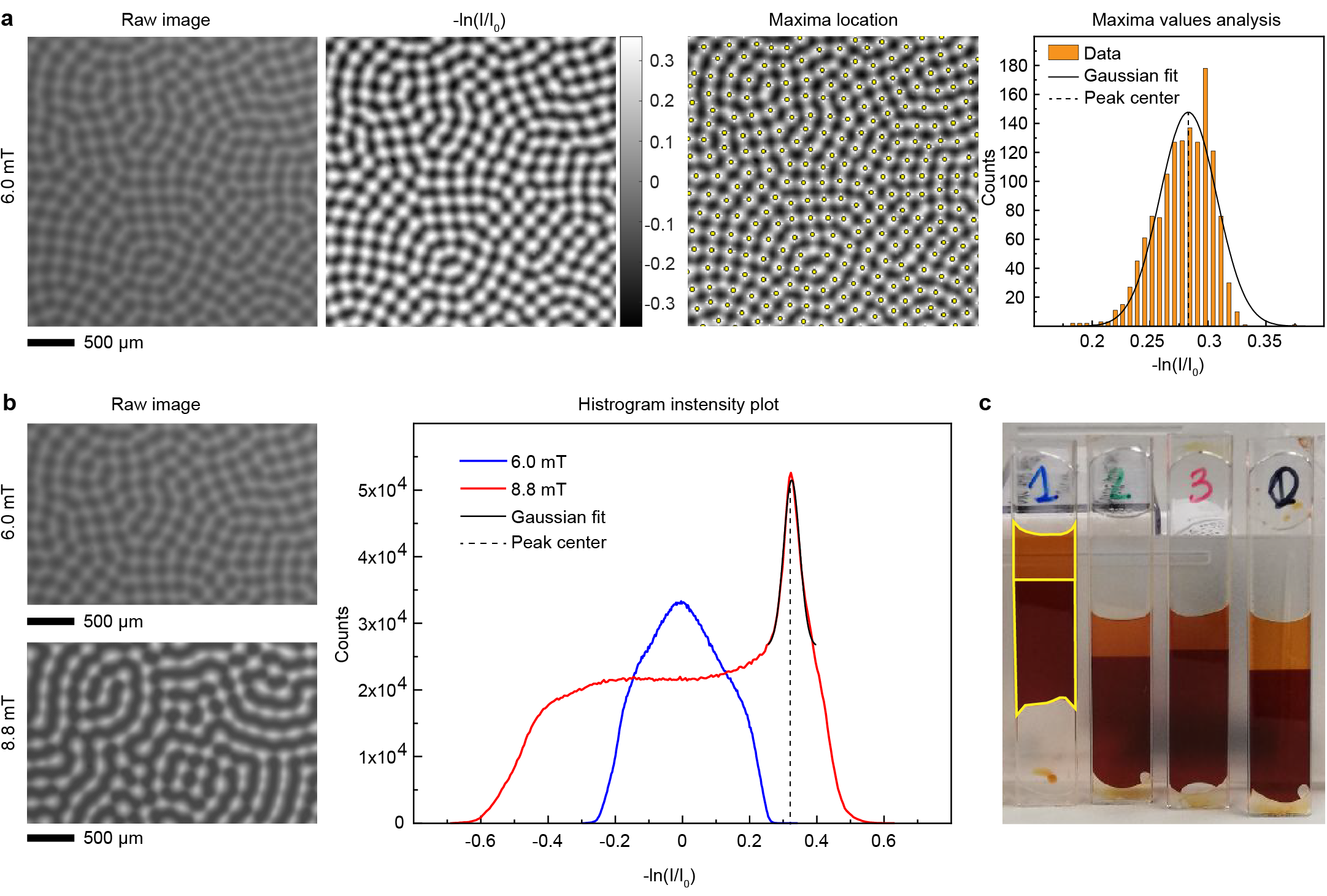}
\label{Extended_Data_Figure.08}
\caption {\textbf{Experimental methods for the amplitude analysis of the instability in the horizontal configuration.} \textbf{a,} From left to right the different steps of the analysis for the amplitude of the instability in the horizontal capillary: raw image, converted image in terms of \(-\ln(I/I_0)\), snapshot of the maxima location tool in the ImageJ software, data relative to the frequency of the intensity values of the maxima and relative Gaussian fit to determine the position of the peak. \textbf{b,} Raw images at two different magnetic fields and histograms of the intensity values for each image. The position of the peak arising when the dense phase touches the upper glass wall is determined with a Gaussian fit. \textbf{c,} Image of the four capillaries used for the study of the instability in the horizontal configuration kept vertical to determine the volume ratio between the two phases. The area is determined by measurement with ImageJ software by drawing a closed spline (see yellow line on the capillary with sample 1). } 
\end{figure}

\begin{figure}[H]
\centering
\includegraphics[width=1\textwidth]{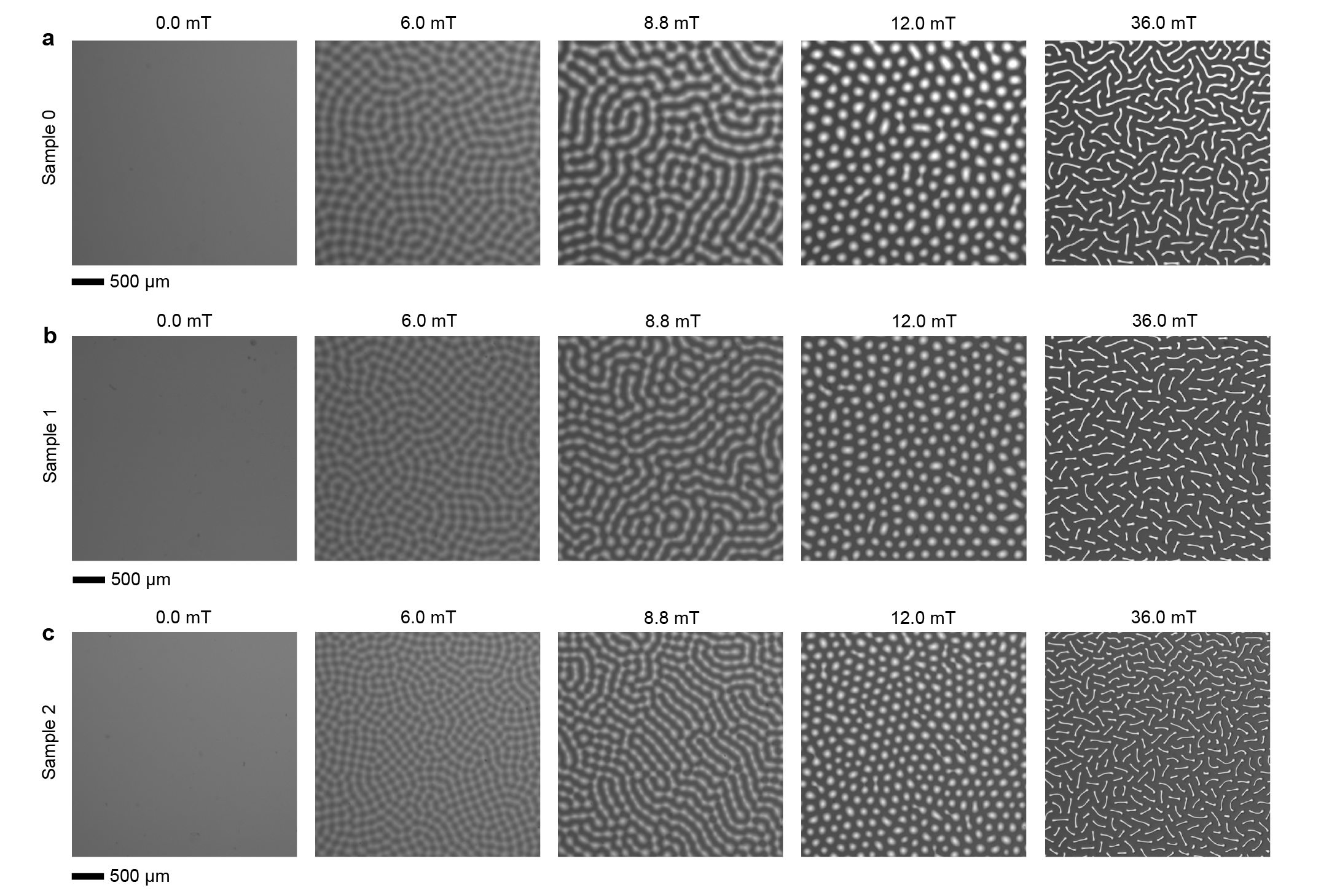}
\label{Extended_Data_Figure.9}
\caption {\textbf{Extended data for the instability in the horizontal configuration.} \textbf{a-c} Microscopy images of the normal field instability in the horizontal capillaries in function of the magnetic field for three different samples. } 
\end{figure}

%% file: 3_SupplementaryNotes.tex
\renewcommand{\thetable}{S\arabic{table}}
\renewcommand{\theequation}{S\arabic{equation}}
\setcounter{table}{0}
\setcounter{equation}{0}

\newpage

\section*{Supplementary Materials}

\subsection*{Supplementary Note S1. Determination of the volumetric composition of the citrated ferrofluid}
The mass fraction of the NPs has been measured letting five samples of 200 \(\upmu\)l of ferrofluid evaporate in oven (Memmert) at 150 degrees for 10h. Weight measurements before and after the evaporation have been performed using the Mettler Toledo balance (0.01 mg precision range). We can calculate the mass fraction of the NPs (\(w_{NP}\)) as:
\begin{equation}
    w_{NP} = \frac{m_{s}}{m_{FF}}
\end{equation}
where \(m_{FF}\) is the ferrofluid mass before evaporation and \(m_{s}\) is the solid residue mass measured after the evaporation. We can also calculate the value of the evaporated water (\(m_{w}\)) as the difference of the other two masses. In the following table the data collected and result of the calculations.

\begin{table}[h]
\centering
\caption{Data used for the calculation of the mass fraction of solid content in the citrated ferrofluid.  }
\begin{tabular}{ |c|c|c|c|c|} 
\hline
Sample number & \(m_{FF}\) (mg) & \(m_{s}\) (mg) & \(m_{w}\) (mg) & \(w_{NP}\) (\%) \\ 
\hline
1 & 249.68 & 63.13 & 186.55 & 25.284 \\
2 & 247.85 & 63.02 & 184.83 & 25.427 \\
3 & 247.29 & 62.81 & 184.48 & 25.399 \\
4 & 247.03 & 62.44 & 184.59 & 25.276 \\
5 & 246.10 & 62.51 & 183.59 & 25.400 \\
\hline
\end{tabular}
\end{table}
\noindent
The resulting average mass fraction of the NPs is \(w_{NP}\) = 25.35 \(\pm\) 0.08 \%. From the measured density value (\(\rho_{FF} =\) 1.28 \(\pm\) 0.08 g/ml) we can calculate the relative volume fraction of the NPs (\(\phi_{NP}\)) as:
\begin{equation}
    \phi_s = \frac{V_s}{V_{FF}} = \frac{V_{FF}-V_w}{V_{FF}} = \frac{m_{FF}/\rho_{FF}-(m_{FF}-m_s/\rho_w)}{m_{FF}/\rho_{FF}} = 1+\frac{\rho_{FF}}{\rho_w} \cdot (w_{NP}-1)
\end{equation}
where \(V_s\) is the volume of the solid content, \(V_{FF}\) is the volume of the ferrofluid, \(V_w\) is the volume of water evaporated and \(\rho_w\) is the density of water (taken as 1.0 g/ml).
The resulting volume fraction is 4.44 \(\pm\) 0.01 \%.

\newpage
\subsection*{Supplementary Note S2. Determination of the mass composition of the FF-ATPS samples}
The main batch of FF-ATPS sample 0 was obtained, following the procedure described in the Methods section. The exact quantities of each material mixed in the preparation are summarized in table S2. 

\begin{table}[h]
\centering
\caption{Weighted quantities of each materials for Sample 0.  }
\begin{tabular}{ |c|c|} 
\hline
Material & \(m_{x}\) (mg)\\ 
\hline
PEG & 740.0 \(\pm\) 0.1 \\
Dextran & 1615.0 \(\pm\) 0.1 \\
Water & 33559.4 \(\pm\) 0.1\\
Ferrofluid & 5112.2 \(\pm\) 0.1\\
\hline
\end{tabular}
\end{table}
\noindent
We can calculate the mass fraction composition of each material in the main sample as \(w_x = m_x/m_{TOT}\), where \(w_x\) and \(m_x\) are the mass fraction and the mass of the material \(x\) and \(m_{TOT}\) is the sum of the masses of all the materials in the dispersion. Remembering that the ferrofluid mass fraction of the NPs is \(w_{NP}\) = 25.35 \(\pm\) 0.08 \% we can calculate how much solid content is present in the main sample and calculate the exact amount of water in it. Note also that the final mass fraction of Dextran has been calculated taking into account that the material is pure at 95\%.

\begin{table}[h]
\centering
\caption{Mass fraction of each materials for Sample 0.  }
\begin{tabular}{ |c|c|} 
\hline
Material & \(w_{x}\) (\%)\\ 
\hline
PEG & 1.8000 \(\pm\) 0.0002\\
Dextran & 3.9284 \(\pm\)0.0002\\
Water & 90.9131 \(\pm\)0.0005\\
Solid from FF & 3.1517 \(\pm\)0.0002\\
\hline
\end{tabular}
\end{table}
\noindent
Note: the missing 0.2 \% from the quantities to reach the 100\% composition are the impurities present in the purchased Dextran.
\newpage
\subsection*{Supplementary Note S3. Magnetic properties of the citrated ferrofluid and FF-ATPS samples.}
For the methodology of the magnetometry measurement see the Methods section and Extended Data Fig. 1c. We can interpolate the magnetization curve data with the Langevin model:
\begin{equation} \label{eu_eqn}
M = nm \left[ \coth \left( \frac{mB}{{k_\mathrm{B}T}}\right) - \left(\frac{mB}{{k_\mathrm{B}T}}\right)^{-1} \right],
\end{equation}
where $M$ is the magnetization of the sample, $B$ is the applied magnetic field, $k_\mathrm{B}$ is the Boltzmann constant, $T$ is the temperature, $n$ is the concentration of the particles (measured in \(m^{-3}\)), and $m$ is the average NP moment (measured in \(A m^2\)). The results of the interpolation for the citrated ferrofluid are: \(m_{\mathrm{FF}} = (1.90 \pm 0.09)\times 10^{-19}\) Am\(^2\) and \(n_{\mathrm{FF}} = (2.0 \pm 0.1)\times 10^{22}\) $m^{-3}$.
From these results we can calculate the saturation magnetization as \(M_{s,FF} = n_{FF} \cdot m_{FF} = 3.8 \pm 0.3 \) kA/m. We can extract the value of the magnetic susceptibility by linear fit of the data near 0 field in the magnetization curve (see inset in Extended Data Fig. 1c). The value obtained in this way is \(\chi_{FF} = 0.0970 \pm 0.0005 \). \\
Following the same approaches used for the citrated ferrofluid we can determine the magnetic properties of each phase of the FF-ATPS samples. In principle, since all the samples are dilutions of the citrated ferrofluid, we should expect that the average magnetic moment of the NPs remains the same. Note that this assumption may be correct only in first approximation, it is possible that NPs of different size partition differently between the light and dense phase so the average moment changes slightly in the two phases. Following this reasoning we fix the average magnetic moment for all the measures as: \( m = (1.90 \pm 0.09)\times 10^{-19}\) Am\(^2\). The resulting properties are summarized in the table below.

\begin{table}[h!]
\centering
\caption{Results for concentration of the NPs ($n$), saturation magnetization (\(M_s\)) and magnetic susceptibility (\(\chi\)) for each phase of samples 0, 1, 2, 3 and for sample 8.}
\begin{tabular}{ |c|c|c|c| } 
 \hline
 Sample phase & $n$ ($m^{-3}$) & \(M_s\) (kA/m) & \(\chi\) \\ 
 \hline
 Sample 0 Light phase & \((8.7 \pm 0.2) \times 10^{20}\) & 0.166 \(\pm\) 0.009 & 0.00208 \(\pm\) 0.00002\\
 Sample 0 Dense phase & \((146 \pm 1)\times 10^{20}\) & 2.8 \(\pm\) 0.1 & 0.0703 \(\pm\) 0.0004\\
  Sample 1 Light phase & \((8.4 \pm 0.2)\times 10^{20}\) & 0.159 \(\pm\) 0.009 & 0.00178 \(\pm\)0.00004\\
 Sample 1 Dense phase & \((146 \pm 1)\times 10^{20}\) & 2.8 \(\pm\)0.1 & 0.0708 \(\pm\)0.0003\\
  Sample 2 Light phase & \((7.9 \pm 0.2)\times 10^{20}\) & 0.151 \(\pm\)0.009 & 0.00196 \(\pm\)0.00005\\
 Sample 2 Dense phase & \((13.3 \pm 0.9)\times 10^{20}\) & 2.5 \(\pm\)0.1 & 0.0639 \(\pm\)0.0003\\
  Sample 3 Light phase & \((11.4 \pm 0.3)\times 10^{20}\) & 0.22 \(\pm\)0.01 & 0.00278 \(\pm\)0.00002\\
 Sample 3 Dense phase & \((133.6 \pm 0.9)\times 10^{20}\) & 2.5 \(\pm\)0.1 & 0.0629 \(\pm\)0.0003\\
 Sample 8 & \((85.4 \pm 0.6)\times 10^{20}\) & 1.62 \(\pm\)0.08 & 0.0399 \(\pm\)0.0002\\
 \hline
\end{tabular}
\vspace*{-0.2in}
\end{table}

\newpage
\subsection*{Supplementary Note S4. Determination of the approximate surface tension from the vertical capillary configuration data.}
To calculate the approximate value of the interfacial tension in the system for the vertical capillary configuration we start from the threshold magnetic field and periodicity data (see Extended Data Fig. 4 for the determination of the threshold magnetic field).

\begin{table}[h]
\centering
\caption{Summary of the analysis for the threshold point of the normal-field instability in the vertical capillaries.}
\begin{tabular}{ |c|c|c|} 
\hline
 & \(B_{c}\) (mT) & \(\lambda_c\) (\(\upmu\)m)\\ 
\hline
Sample 0 & 7.6\(\pm\) 0.9 & 135.7 \(\pm\) 0.5  \\
Sample 1 & 6 \(\pm\) 1 & 118.2 \(\pm\) 0.5\\
Sample 2 & 5.3 \(\pm\)0.9 & 86.5 \(\pm\) 0.5\\
\hline
\end{tabular}
\label{2.5instabdata}
\end{table}
\noindent
The second step is to convert the values of threshold magnetic field into critical magnetization using the following relation:
\begin{equation}
    M (A/m) = \mu_0 \chi B (T) 
\end{equation}
Note that this relationship is valid only if the values of magnetic field are small in relationship to the point at which the magnetization begins to saturate in the magnetization curve. In our case all the samples reach the critical magnetization at very small fields when still in the linear part of the magnetization curve (see Extended Data Fig. 2). In our case the value \(\chi\) must be taken as the difference between the magnetic susceptibility of the dense and light phase (see Table S4). Once we calculated the critical magnetization data we can use eq. 6 to calculate the surface tension remembering that \(k_c = 2\pi/\lambda_c\) and that the thickness of the capillary \(h\) in these experiments was 0.10 \(\pm\) 0.01 mm. 

\begin{table}[h]
\centering
\caption{Results of the calculations for the determination of the surface tension from the normal-field instability in the vertical capillaries.}
\begin{tabular}{ |c|c|c|c|c|c|c|} 
\hline
 & \(\chi\) & \(\Delta\rho\) (g/ml) & \(M_{c}\) (kA/m) & \(k_ch\) & \(f(k_ch)\) & \(\gamma\) (\(\upmu\)N/m)\\ 
\hline
Sample 0 & 0.0666\(\pm\) 0.0005 & 0.06 \(\pm\) 0.01 & 0.40 \(\pm\) 0.05 & 4.6 \(\pm\) 0.5 & 4.8 \(\pm\) 0.5 & 1.2 \(\pm\) 0.2 \\
Sample 1 & 0.0675\(\pm\) 0.0005 & 0.05 \(\pm\) 0.01 & 0.32 \(\pm\) 0.06 & 5.3 \(\pm\) 0.5 & 5.8 \(\pm\) 0.6 & 0.7 \(\pm\) 0.1\\
Sample 2 & 0.0605\(\pm\) 0.0005 & 0.05 \(\pm\) 0.01 & 0.25 \(\pm\) 0.05  & 7.3 \(\pm\) 0.7 & 8.8 \(\pm\) 0.9 &  0.33 \(\pm\) 0.07 \\
\hline
\end{tabular}
\label{2.5instabresult}
\end{table}
\noindent
 Note: the value of \(\Delta\rho\) for each sample has been calculated from the density values for the two phases in Table 3.
\newpage
\subsection*{Supplementary Note S5. Determination of the approximate surface tension and critical magnetic field from the horizontal capillary configuration data.}
To calculate the approximate value of the interfacial tension in the system and the theoretical value of the threshold magnetic field \(B_c\) for the horizontal capillary configuration we start from  periodicity data. 

\begin{table}[h]
\centering
\caption{Summary of the analysis for the critical point of the normal-field instability in the horizontal capillaries.}
\begin{tabular}{ |c|c|} 
\hline
 &  \(\lambda_c\) (\(\upmu\)m)\\ 
\hline
Sample 0  & 244.4 \(\pm\) 0.8  \\
Sample 1  & 225.8 \(\pm\) 0.8\\
Sample 2  & 189.4 \(\pm\) 0.7\\
\hline
\end{tabular}
\label{3instabdata}
\end{table}
\noindent
Using eq. 1 and 2 we can calculate an approximate value of the interfacial tension and threshold magnetic field (from the calculated value of critical magnetization obtained from eq. 2 and the conversion with eq. S4). The last value missing to perform the calculations of eq. 2 is the magnetic permeability \(\mu\) in both the light and dense phase for each sample but it can be calculated from the values of \(\chi\) in table S4 as: \(\mu = \mu_0 (1+\chi)\).

\begin{table}[h]
\centering
\caption{Results of the calculations for the determination of the surface tension and threshold magnetic field from the normal-field instability in the horizontal capillaries.}
\begin{tabular}{ |c|c|c|c|c|c|} 
\hline
 & \(\chi\) & \(\Delta\rho\) (g/ml) & \(\gamma\) (\(\upmu\)N/m) & \(M_{c}\) (kA/m) & \(B_c\) (mT) \\ 
\hline
Sample 0 & 0.0666\(\pm\) 0.0005 & 0.06 \(\pm\) 0.01 & 0.9 \(\pm\) 0.1 & 0.27 \(\pm\) 0.03 & 5.0 \(\pm\) 0.6\\
Sample 1 & 0.0675\(\pm\) 0.0005 & 0.05 \(\pm\) 0.01 & 0.6 \(\pm\) 0.1 & 0.23 \(\pm\) 0.03 & 4.3 \(\pm\) 0.6 \\
Sample 2 & 0.0605\(\pm\) 0.0005 & 0.05 \(\pm\) 0.01 & 0.45 \(\pm\) 0.09  & 0.21 \(\pm\) 0.03 & 4.4 \(\pm\) 0.6  \\
\hline
\end{tabular}
\label{3instabresult}
\end{table}

\newpage
\subsection*{Supplementary Note S6. Derivation of eq. 3 and 4 from the Beer-Lambert law.}
The Beer-Lambert law relates the attenuation of the light intensity passing through a material to its properties. In particular the attenuation coefficient can be written as:
\begin{equation}
    A = \epsilon h n
\end{equation}
where \(A\) is the attenuation coefficient, \(\epsilon\) is the molar attenuation coefficient of the material, \(h\) is the thickness of the material and \(n\) is its concentration. Starting from this, we can write the attenuation of light as:
 \begin{equation}
      \delta I = - I \epsilon  \delta z   n
 \end{equation}
where $z$ is the path length and $I$ is the light intensity. Integrating the equation over the thickness of the layer, we get:
 \begin{equation}
    \int_{I_{\mathrm{in}}}^{I_{\mathrm{out}}} \frac{dI}{I} = \int_{0}^{h}  - \epsilon n dz,
 \end{equation}
therefore,
 \begin{equation}\label{integralI}
\ln \frac{I_{\mathrm{out}}}{I_{\mathrm{in}}} = - \epsilon  h n.
 \end{equation}
where \(I_{\mathrm{in}}\) and \(I_{\mathrm{out}}\) are the intensity before and after passing through the material respectively. The light intensity outside of the glass can be calculated as
 \begin{equation}{\label{eq:Iout}}
     I_{\mathrm{out}}=I_{\mathrm{in}}\exp({- \epsilon n h}),
 \end{equation}
The average concentration can be expressed as:
\begin{equation} \label{beerconc}
    n = -\frac{1}{\epsilon h}\left(\ln \frac{I_{\mathrm{out}}}{I_{\mathrm{in}}}\right).
\end{equation}
In the same way we can estimate the thickness of a layer as:
\begin{equation} \label{beerthick}
    h = -\frac{1}{\epsilon n}\left(\ln \frac{I_{\mathrm{out}}}{I_{\mathrm{in}}}\right).
\end{equation}
In general, when comparing two samples of the same material with different concentration, we can completely avoid to measure the value of \(I_{\mathrm{in}}\). In fact, for two layers of equal thickness, we can write:
 \begin{equation}
 \Delta n = n_1 - n_2 = -\frac{1}{\epsilon h}\left(\ln \frac{I_{\mathrm{out, 1}}}{I_{\mathrm{in, 1}}} - \ln \frac{I_{\mathrm{out,2}}}{I_{\mathrm{in, 2}}} \right) = -\frac{1}{\epsilon h} \left(\ln \frac{I_{\mathrm{out, 1}}}{I_{\mathrm{out, 2}}} \right)
 \end{equation}
\noindent 
Therefore we have obtained the desired shape of the Beer-Lambert law for eq. 3. \\
Another useful case for which we can derive the formula is if the thickness of our sample changes as a function of a spatial coordinate. In this case we can derive the variation of thickness of the layer as a function of the spatial coordinate \(x\) from eq. S11 following the same approach as for eq. S12 as:
 \begin{equation}
\Delta h (x) = h_0 - h(x) = -\frac{1}{\epsilon n} \left(\ln \frac{I_{\mathrm{out}}(x)}{I_{\mathrm{out,0}}} \right)
 \end{equation}
where \(I_{\mathrm{out,0}}\) is the intensity in a region of known thickness and \(I_{\mathrm{out}}(x)\) is the output intensity in the region we want to measure. We can further develop this formula also in the case of two layers that change thickness in function of the spatial coordinate if the sum of the thickness of the two layers is constant (\(h_{1+2} = h_1 + h_2\)). 
Following the same reasoning as for the single layer case we can obtain the equivalent of eq. S8 for the case of two layers as:
\begin{equation} \label{twolayers}
    \alpha_1 h_1 + \alpha_2 h_2 = -\ln \left(\frac{I_{\mathrm{out}}}{I_{\mathrm{in}}}\right)
\end{equation}
where \(\alpha_{1,2}\) are parameters that take into account both the molar attenuation coefficient and the concentration of each material. We can rearrange eq. S14 as: 
\begin{equation}
    h_1 = \frac{1}{\alpha_1-\alpha_2} \left[\ln\left( \frac{I_{\mathrm{in}}}{I_{\mathrm{out}}}\right) - h_{1+2}\alpha_2\right]
\end{equation}
At this point we can generalize the formula to determine the thickness of the first layer in a random point of the sample as:
\begin{equation}
    h(x) = \frac{1}{\alpha_1-\alpha_2} \left[\ln\left( \frac{I_{\mathrm{in}}}{I_{\mathrm{out}}(x)}\right) - h_{1+2}\alpha_2\right]
\end{equation}
And write again the relation for the variation of thickness along the direction \(x\) as:
\begin{equation}
    \Delta h(x) = h(x) - h_0 = -\frac{1}{\alpha} \ln\left( \frac{I_{\mathrm{out}}(x)}{I_{\mathrm{out,0}}}\right)
\end{equation}
where \(\alpha = \alpha_1-\alpha_2\) and \(I_{\mathrm{out,0}}\) is the output intensity in a region where the thickness of the layer (\(h_0\)) is known. Therefore, we have obtained a general formula for eq. 4. Note that this formula can be applied both to the case of a layer changing thickness in function of a spatial coordinate and in the case of a single point in the interface changing its thickness in function of the magnetic field.